\documentclass[twocolumn]{aastex631}
\usepackage{graphicx}
\usepackage{amssymb}
\usepackage[fleqn]{amsmath}
\usepackage{verbatim}
\usepackage{natbib} 
\usepackage[utf8]{inputenc}
\usepackage{CJK}
\usepackage{xcolor}

\begin{document}

\defcitealias{wold22}{W22}

\title{UNCOVERing the Faint-End of the $z\sim7$ [OIII] Luminosity Function with {\it JWST}'s F410M Medium Bandpass Filter}

\author[0000-0002-0784-1852]{Isak G. B. Wold}
\affil{Astrophysics Science Division, Goddard Space Flight Center, Greenbelt, MD 20771, USA}
\affil{Department of Physics, The Catholic University of America, Washington, DC 20064, USA }
\affil{Center for Research and Exploration in Space Science and Technology, NASA/GSFC, Greenbelt, MD 20771}

\author[0000-0002-9226-5350]{Sangeeta Malhotra} 
\affiliation{Astrophysics Science Division, Goddard Space Flight Center, Greenbelt, MD 20771, USA}

\author[0000-0002-1501-454X]{James E. Rhoads} 
\affiliation{Astrophysics Science Division, Goddard Space Flight Center, Greenbelt, MD 20771, USA}

\author[0000-0003-1614-196X]{John R. Weaver}
\affiliation{Department of Astronomy, University of Massachusetts Amherst, Amherst MA 01003, USA}

\author[0000-0001-9269-5046]{Bingjie Wang (\begin{CJK*}{UTF8}{gbsn}王冰洁\ignorespacesafterend\end{CJK*})}
\affiliation{Department of Astronomy \& Astrophysics, The Pennsylvania State University, University Park, PA 16802, USA}
\affiliation{Institute for Computational \& Data Sciences, The Pennsylvania State University, University Park, PA 16802, USA}
\affiliation{Institute for Gravitation and the Cosmos, The Pennsylvania State University, University Park, PA 16802, USA}

\begin{abstract}
Strong emission from doubly ionized oxygen is a beacon for some of the most intensely star forming galaxies known.  {\it JWST} enables the search for this beacon in the early universe with unprecedented sensitivity.  In this work, we extend the study of faint [OIII]$_{5008}$ selected galaxies by an order of magnitude in line luminosity.
We use publicly available UNCOVER DR1 {\it JWST}/NIRCam and {\it HST} imaging data of the cluster lensing field, Abell 2744, to identify strong (rest-frame EW$>500$\AA) [OIII]$_{5008}$ emitters at $z\sim7$ based on excess F410M flux.  We find $N=68$ $z\sim7$ [OIII] candidates, with a subset of $N=33$ that have deep {\it HST} coverage required to rule-out lower redshift interlopers (13.68 arcmin$^2$ with F814W $5\sigma$ depth $>28$ AB).  Such strong emission lines can lead to very red colors that could be misinterpreted as evidence for old, massive stellar populations, but are shown to be due to emission lines where we have spectra.  
Using this deep {\it HST} sample and completeness simulations, which calculate the effective survey volume of the UNCOVER lensing field as a function of [OIII] luminosity, we derive a new [OIII] luminosity function (LF) extending to $41.09<\rm{log}_{10}(L/\rm{erg\,s}^{-1})<42.35$ which is an order of magnitude deeper than previous $z\sim6$ [OIII] LFs based on {\it JWST} slitless spectroscopy.  This  LF is well fit by a power law with a faint-end slope of $\alpha=-2.07^{+0.22}_{-0.23}$.  
There is little or no evolution between this LF and published [OIII] LFs at redshifts $3\lesssim z\lesssim7$,and no evidence of a turnover at faint luminosities. The sizes of these extreme [OIII] emitters are broadly similar to their low redshift counterparts, the green peas. The luminosity function of [OIII] emitters matches that of Lyman-$\alpha$ at the bright end, suggesting that many of them should be Lyman-$\alpha$ emitters.
\end{abstract}

\section{Introduction}

{\it Spitzer}/IRAC studies have identified a population of extreme emission line galaxies at $z>6$ that significantly perturb [3.6]$-$[4.5] $\mu$m broad-band colors due to high equivalent width [OIII]+H$\beta$ lines \citep[e.g.][]{egami05,chary05,Schaerer09,stark13,Labbe13,Smit14,Roberts-Borsani16,endsley21}. These strong emission lines produce red colors that need to be properly accounted for \citep[e.g.,][]{mclinden11, cowie11} to avoid misinterpretation as evidence of old, massive stellar populations. A recent IRAC study \citep{debarros19} suggests that the luminosity density of these extreme emitters increases by a factor of $\sim5$ from a redshift of $z=2$ to $z=8$.  This is in contrast to the evolution of the cosmic star formation rate density which declines by an order of magnitude over the same redshift range \citep[e.g.,][]{bouwens15,finkelstein15}, suggesting that at a given SFR the production of ionizing flux - needed for strong [OIII] emission - increases with redshift.  This increasing strength and the overall prevalence of [OIII]+H$\beta$ emission at $z>6$ emphasizes its potential to efficiently identify galaxies at the highest redshifts. 

With the arrival of {\it JWST}, NIRCam wide-field slitless spectroscopy (WFSS) studies \citep{sun23,matthee23} have confirmed a large population of strong [OIII] emitters at $z\sim6$, albeit with slightly smaller number densities at intermediate luminosities ($10^{42}$ erg s$^{-1}$) when compared to the recent IRAC $z\sim8$ study \citep{debarros19}, indicating $z=6$ to $8$ redshift evolution or slight tension with the previous study.  Regardless, these {\it JWST} studies spectroscopically confirm the ubiquity of strong [OIII] line emitters and further highlight their ability to trace the galaxy density in the Epoch of Reionization

In this paper, we demonstrate the utility of {\it JWST} medium bandpasses to find extreme high-redshift emitters \citep[also see][]{Withers23,Rinaldi23,suess24,llerena24}.  Specifically, we search for F410M-excess objects to identify $z\sim7$ [OIII] emitters with rest-frame equivalent widths (EWs) above $500$\AA. We show that this is an efficient method of identifying high-redshift sources that complements (1) broadband (BB) surveys that require a significant continuum detection, and (2) slitless spectroscopic surveys that typically contend with higher background levels and overlapping spectra.

For our F410M-based search, we use the publically available Ultradeep NIRSpec and NIRCam ObserVations before the Epoch of Reionization (UNCOVER) survey \citep{bezanson22,fujimoto23,weaver24,wang24,suess24}.  This is a Cycle 1 {\it JWST} Treasury program, which includes ultradeep ($\sim30$AB) imaging with a $\sim30$ arcmin$^2$ field of view around the Abell 2744 galaxy cluster at $z= 0.308$.  The cluster has copious {\it HST} data and has a well-characterized lensing model \citep{furtak23} facilitating our search for $z\sim7$ [OIII] emitters.

 Simulations have suggested that the escape of ionizing flux is easier for low-mass galaxies where gas-covering fractions are lower and starbursts are able to clear out channels in the interstellar medium \citep[e.g.,][]{anderson17, finkelstein19}.  Furthermore, the observed steepening of the faint-end slope of the UV luminosity function with increasing redshift indicates that faint, low-mass galaxies have the numbers to produce the ionizing flux needed for reionization \citep[e.g.,][]{finkelstein15,bouwens22}.  Thus, low-mass, star-forming galaxies have emerged as one of the most likely drivers of reionization.  Our selected extreme EW objects are selected regardless of their continuum brightness, allowing us to probe low-mass galaxies.  Additionally, the UNCOVER field has a strong gravitational lensing cluster that enables us to detect even lower-mass extreme emitters that are in the midst of formation.

As lower masses are probed, correspondingly lower metallicities are expected \citep[e.g.,][]{tremonti04, andrews13}.  If this metal decline is sufficient to deplete oxygen \citep[e.g.,][]{kojima20}, we may see a deficit in the number density of faint [OIII] emitters which could cause a flattening or a turnover in the faint-end slope of the luminosity function.  We look for this signature with our F410M-excess selection which goes $\sim1$ dex deeper than past surveys, giving us [OIII] number density constraints at luminosities of $\gtrsim10^{41}$ erg s$^{-1}$.

Throughout this work, all EWs are
rest-frame, [OIII] refers to the [OIII]$_{5008}$ line while [OIII]+H$\beta$ refers to the [OIII]$_{5008,4960}$ and H$\beta$ lines.  All magnitudes are in the AB magnitude system ($m_{\mbox{\footnotesize{AB}}}=31.4-2.5$log$_{10}f_{\nu}$
with $f_{\nu}$ in units of nJy). Distances in Mpc units and volumes in Mpc$^3$ units are comoving. We adopt a flat $\Lambda$CDM cosmology
with $\Omega_{m}=0.3$, $\Omega_{\Lambda}=0.7$, and H$_{0}=70$ km
s$^{-1}$Mpc$^{-1}$.

\begin{figure}
\begin{centering}
\includegraphics[width=8.5cm]{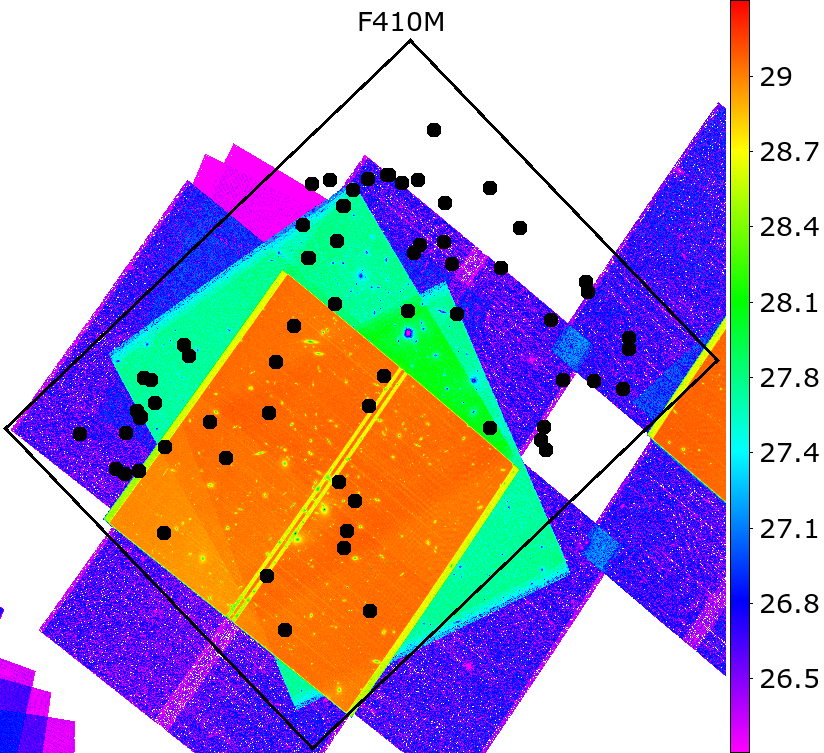}\caption{$5\sigma$ depth map for the {\it HST}-ACS F814W mosaic compared to the $4.7\times6.1'$ coverage of the F410M mosaic.  Filled black circles show the locations of our candidate [OIII] emitters at $z\sim7$. The other {\it HST}-ACS mosaics - which are our main veto-bands - have very similar exposure maps and cover $85\%$ of the F410M field-of-view.  All {\it JWST}-NIRCam mosaics encompass the F410M mosaic except F090W which intersects the western corner, covering only $2$ arcmin$^2$ of the F410M field-of-view.  The displayed $5\sigma$ depths are $0.3''$ diameter aperture depths, and we list median depths for all bands in Table \ref{dtable}.   
}\label{fov}
\end{centering}
\end{figure}

\begin{deluxetable}{lcc} 
\tablecolumns{3} 
\tablewidth{0pc} 
\tablecaption{Uncover's median $5\sigma$ depth measured in $0.3''$ diameter apertures.\label{dtable}} 
\tablehead{ 
\colhead{Instrument\phm{adfad}} & \colhead{Filter} & \colhead{$5\sigma$ Depth}}
\startdata 
HST-ACS & F435W & 28.9 \\
HST-ACS & F606W & 29.1 \\
HST-ACS & F814W & 29.1 \\
JWST-NIRCam & F090W & 28.7 \\
HST-WFC3 & F105W & 28.5 \\
JWST-NIRCam & F115W & 28.6 \\
HST-WFC3 & F125W & 28.2 \\
HST-WFC3 & F140W & 28.8 \\
JWST-NIRCam & F150W & 28.7 \\
HST-WFC3 & F160W & 28.2 \\
JWST-NIRCam & F200W & 28.7 \\
JWST-NIRCam & F277W & 29.2 \\
JWST-NIRCam & F356W & 29.4 \\
JWST-NIRCam & F410M & 28.9 \\
JWST-NIRCam & F444W & 29.1
\enddata  

\end{deluxetable}

\section{Observations}

We downloaded the Abell 2744 mosaics available from the UNCOVER DR1\footnote{for the latest data release see: \url{https://jwst-uncover.github.io/}} \citep{bezanson22}.  This includes {\it JWST} data from GO-2561 (UNCOVER), ERS-1324 (GLASS), and DD-2767; along with {\it HST} data from programs 11689 (PI: Dupke), 13386 (PI: Rodney), 13495 (PI: Lotz), 13389 (PI: Siana), 15117 (PI: Steinhardt), and 17231 (PI: Treu). We summarize bandpasses and median $5\sigma$ depths of the mosaics that we use for candidate selection and measurement of photometric redshifts in Table \ref{dtable}.  These are the same bandpasses utilized by the UNCOVER DR1 catalog paper \citep{weaver24}. As expected, their median $5\sigma$ depths measured in $0.32''$ diameter apertures (reported in their Table 1) are found to be in good agreement with our $0.3''$ diameter aperture depth measurements.  We measure image depth by computing the flux spread in randomly sampled sky apertures within the F410M field-of-view.  The intersection of the F410M and the F090W image has a small $\sim2$ arcmin$^{2}$ area. For this reason, we measure the F090W depth over its full field of view ($\sim13$ arcmin$^2$). 

Our primary bandpass is F410M which has a mosaic covering $28.7$ arcmin$^2$.  The other {\it JWST}-NIRCam mosaics encompass the F410M footprint except the F090W image which only intersects $7\%$ of the F410M mosaic's area.  All {\it HST}-ACS mosaics (F814W, F606W, and F435W) have a similar exposure map, and in Figure \ref{fov} we demonstrate this by showing the F814W mosaic's footprint which covers $85\%$ of the F410M mosaic.  The displayed depth map was created by scaling the inverse of the square root of the UNCOVER DR1 weight map to our $0.3''$ diameter aperture depth measurement. Referring to Figure \ref{fov}, the {\it HST}-WCF3 mosaics that overlap with F410M coverage are centered on the deepest ($5\sigma=29$ mag) F814W pointing but with a smaller $5$ to $19$ arcmin$^{2}$ field of view.  These WCF3 bandpasses (F105W, F125W, F140W, F160W) are not used in our $z\sim7$ [OIII] selection but are used to constrain photometric redshifts over the area with coverage. 

\begin{figure}
\begin{centering}
\includegraphics[width=8.5cm]{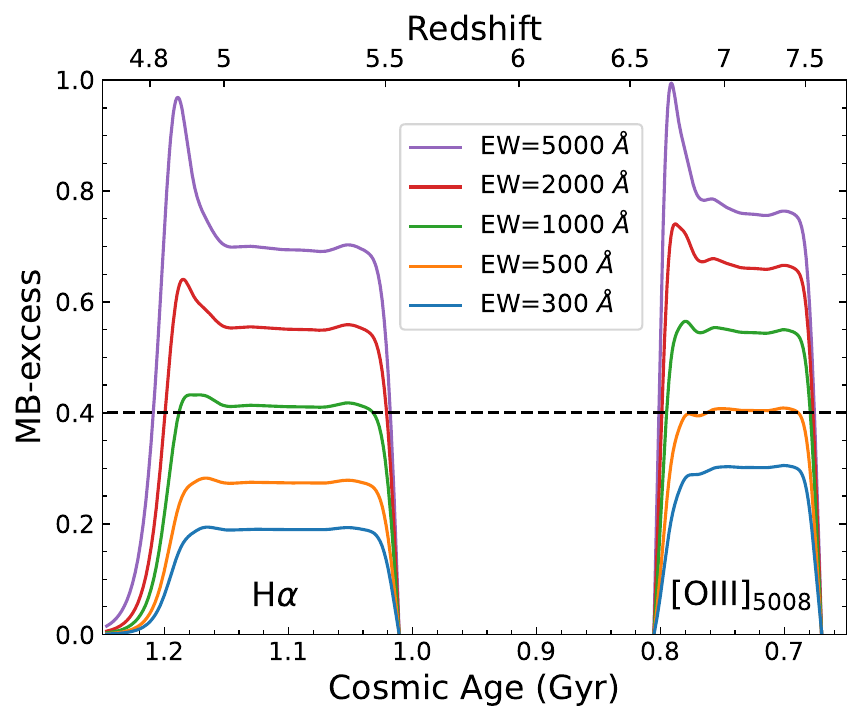}\caption{Expected F444W$-$F410M color for $z\sim7$ [OIII] emitters and $z\sim5$ H$\alpha$ emitters. Our candidates must fall above the F444W$-$F410M$=0.4$ dashed horizontal line.  A high MB-excess peak of $\sim1$ mag is predicted for EW$=5000$\AA\ emitters because the tail of the F410M response goes slightly bluer than the F444W response resulting in a narrow wavelength range where emission is only influencing the F410M bandpass. 
}\label{color_select}
\par\end{centering}
\end{figure}

\begin{figure*}
\begin{centering}
\includegraphics[height=6.5cm]{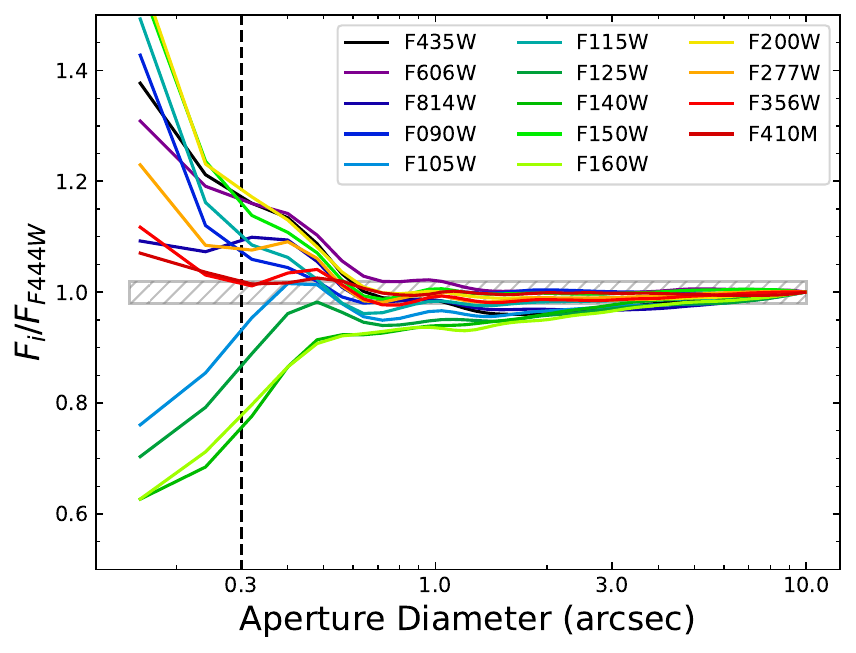}\includegraphics[height=6.5cm]{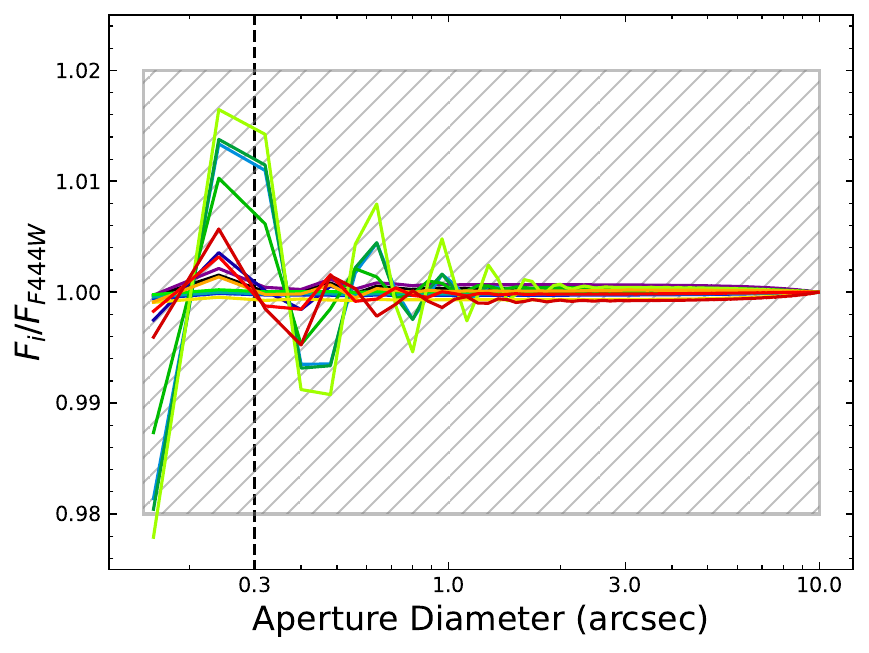}
\caption{PSF curves of growth normalized by the F444W COG before and after PSF-matching. In both panels the hatched region shows where the PSF profiles relative to the F444W PSF differ by $2\%$ or less.  We select $z\sim7$ [OIII] emitters based on aperture photometry with a diameter of $D=0.3''$ (vertical dashed line).}\label{psf_match}
\par\end{centering}
\end{figure*}  

\section{[OIII] Candidates at $z\sim7$}\label{select0}

\subsection{Minimum Equivalent Width}\label{ewselect}
We isolate a sample of $z\sim7$ [OIII] emitters by looking for medium-bandpass (MB) excess galaxies, where the excess is measured by the F444W$-$F410M color.  To help define our MB-excess cut, we assumed a spectral shape for our objects of interest and then used the F444W and F410M response functions\footnote{based on the NIRCam detector averaged response: \url{https://jwst-docs.stsci.edu/jwst-near-infrared-camera/nircam-instrumentation/nircam-filters}} to solve for the expected MB-excess as a function of redshift and EW. We carried out this procedure for $z\sim7$ [OIII] emitters and for $z\sim5$ H$\alpha$ emitters which are likely to be our primary ``low'' redshift interlopers.   For the spectral shape, we use a flat $f_{\nu}$ continuum with unresolved emission lines.  We approximate the emission line flux by assuming the following line ratios: [OIII]$_{5008}/$[OIII]$_{4960}=3$, 
 [OIII]$_{5008}/$H$\beta=6.72$, and H$\alpha/$H$\beta=2.86$.  The [OIII]$_{5008}/$H$\beta$ line ratio is consistent with recent {\it JWST}/NIRCam WFSS studies of $z\sim6$ [OIII] emitters, where \citet{sun23} found a median ratio of $6.72$ and \citet{matthee23} found an average ratio of $6.3$.   We conduct two runs where we either alter the H$\alpha$ or the [OIII]$_{5008}$ EW and show the results in Figure \ref{color_select}.

 We find that a color cut of F444W$-$F410M $>0.4$ selects $z\sim7$ [OIII] emitters with EW $>500$\AA, while only selecting relatively extreme EW $>1000$\AA\  H$\alpha$ emitters at $z\sim5$.  Assuming a $z\sim5$ H$\alpha$ EW scale length of $500$\AA\ \citep[e.g.,][]{faisst16}, an EW $>1000$\AA\ cut will remove $\sim85\%$ of the H$\alpha$ population. We use the deep F814W imaging   -- $0.3''$ aperture diameter $5\sigma$ AB depths of $28\pm1$ -- available for $85\%$ of the F410M field-of-view to help identify any remaining H$\alpha$ emitters.

 \subsection{Background Subtraction}
We adopt a fast and easy-to-implement \textsc{Source Extractor} \citep{bertin96} background subtraction procedure and then verify that our derived photometric redshifts are consistent with those cataloged in \citet{weaver24}. Bright cluster galaxies and intra-cluster light produce a wavelength-dependent foreground that may bias the color measurements of our $z\sim7$ objects of interest.  However, our [OIII] candidates are primarily selected based on their F444W$-$F410M color.  These two bandpasses have very similar effective wavelengths which helps to mitigate the effect of wavelength-dependent foregrounds.

We perform background subtraction with \texttt{BACK\_SIZE} $=32$ pixels and \texttt{BACK\_FILTERSIZE} $=4\times4$, which set the \textsc{Source Extractor} background mesh size and the number of meshes used when median smoothing the background image, respectively.  Before this background subtraction step, we $2\times2$ binned the {\it JWST} short-wavelength (SW) mosaics' $0.02''$ pixels to match the long-wavelength  (LW) and {\it HST} images' $0.04''$ pixel size and pixel grid.  All subsequent analyses use these re-binned F090W, F105W, F150W, and F200W SW mosaics. A uniform pixel grid for all bandpasses is needed for our \textsc{Source Extractor} procedure which uses double-image mode.

The UNCOVER catalog paper adopts the background subtraction procedure described in \citet{ferrarese06}. Here, each bright cluster galaxy's light profile is modeled. This fit profile is then subtracted from the science image, removing the cluster foreground in each bandpass.  As discussed in Section \ref{photoz}, we find that our photometric redshifts are in close agreement with \citet{weaver24}'s results, suggesting that our simple approach can remove foreground cluster light sufficiently.

\begin{figure}
\begin{centering}
\includegraphics[width=8.5cm]{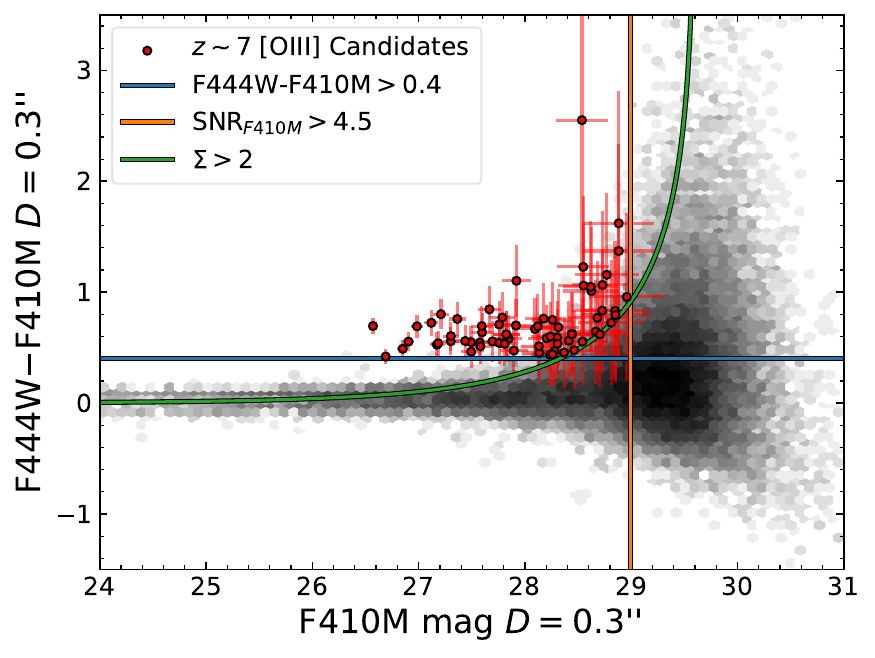}
\caption{F410M-excess selection for the UNCOVER field. The overall source distribution is shown by a log$_{10}$-scaled 2D grey density histogram with selected $z\sim7$ [OIII] candidates shown by red points. The vertical orange line indicates the median $4.5\sigma$ depth for the UNCOVER MB image.  The horizontal blue line indicates the $0.4$ MB-excess cut used to select EW $>500$\AA\ [OIII] candidates.  The green curve shows the median $2\Sigma$ cut (see Equations \ref{seleq} \& \ref{sigeq}). }\label{uncover_select}
\par\end{centering}
\end{figure}

\subsection{PSF-matching}
To ensure accurate color measurements with fixed aperture photometry, we PSF-match all images to the F444W image which has the largest PSF.  We determine the PSFs for all bandpasses by isolating non-saturated stars and then performing a weighted summation.  Stars were identified by selecting objects within the stellar plume of the half-light radius versus magnitude plot.  This stellar plume is occupied by unresolved stars that maintain a constant radius regardless of their brightness. We created $10\times10''$ postage stamps for each star and then performed a weighted summation.  Neighboring sources to our stars were identified with a \textsc{Source Extractor} segmentation map and were given zero weight when making the stacked PSF image.

With 15 PSF postage stamp images in hand -- one per bandpass -- we made smoothing kernels using \textsc{Pypher} \citep{boucaud16}. \textsc{Pypher} is based on Wiener filtering and can account for anisotropic features in the PSF.  In Figure \ref{psf_match}, we show the PSF growth curve relative to the F444W PSF growth curve before and after PSF-matching \citep[see the UNCOVER catalog paper,][for similar results]{weaver24}.  We find that all matched PSFs have growth curves that agree at the $2\%$ level or better.  With these PSF-matched images, we used \textsc{Source Extractor} in double-image mode to create an F410M-selected catalog. We used the non-PSF-matched F410M mosaic as the detection image and the 15 PSF-matched mosaics as the measurement images.

\subsection{Bright Star and Edge Mask}

Given UNCOVER's relatively small field of view, we manually mask bright stars and the edge of the F410M mosaic.  The outer $\sim18''$ image border has reduced exposure times for the available {\it JWST} mosaics.  Our initial [OIII] target selection found excessive artifacts within this region, resulting in our decision to mask this area.  The resulting edge plus bright star mask reduces the F410M survey area from $28.7$ to $20.7$ arcmin$^{2}$.

\subsection{Candidate Selection}\label{select} 

We use the F444W$-$F410M color and the redshifted Ly$\alpha$ break at $(1+z)1216$\AA\ to isolate $z\sim7$ [OIII] emitters.  Bandpasses F606W and F435W fall completely below the $(1+z)1216$\AA\ spectral break, and here we require $<3\sigma$ non-detections.  Bandpass F814W spans $0.70$ to $0.96 \mu$m and may contain continuum and line flux redward of this break for [OIII] emitters at the low end of our redshift range, $6.7<z<7.5$.  For this reason, selected [OIII] candidates are allowed to have F814W counterparts but this counterpart must be faint relative to the continuum, F814W-F115W $>1$.

For the UNCOVER [OIII] selection (see Figure \ref{uncover_select}), we use $0.3''$ diameter apertures on images PSF-matched to the F444W image images via:
\begin{footnotesize}
\begin{equation}\label{seleq}
\begin{array}{ll}
&{[(F444W-{\rm{F410M}>0.4}\;\&\;{\rm{SNR}_{\rm{F444W}}>3})\;{\rm{or}\;{SNR}_{\rm{F444W}}<3}]}\;\& \\
&{[(F814W-{\rm{F115W}>1.0}\;\&\;{\rm{SNR}_{\rm{F814W}}>3})\;{\rm{or}\;{SNR}_{\rm{F814W}}<3}]}\;\& \\
& {\rm{ SNR}_{\rm{F606W}}<3}\;\&\ {\rm{ SNR}_{\rm{F435W}}<3}\;\&\  {\rm{SNR}_{\rm{F410M}}>4.5}\;\&\ \Sigma>2
\end{array}
\end{equation}
\end{footnotesize}

\noindent where
\begin{equation}\label{sigeq}
\Sigma=\frac{f_{\rm{F410M}}-f_{F444W}}{\sqrt{\sigma^{2}_{\rm{F410M}}+\sigma^{2}_{F444W}}}
\end{equation}

This selection isolated $95$ candidates within the field.  We manually inspected all candidates for potential problems (e.g., diffraction spikes, cosmic rays, visual veto-band counterparts below our formal $3\sigma$ cut) by viewing postage stamp cutouts over all the available bandpasses.  We find a final sample of $N=68$ [OIII] candidates at $z\sim7$. $N=33$ objects have deep veto coverage (F814W $5\sigma$ depth $>28$ AB), while $N=13$ candidates have no {\it HST}-ASC coverage (see Figure \ref{fov}).  These $13$ candidates are more likely to be contaminated by foreground objects, and we designate the 33 objects with deep {\it HST}-ACS coverage as our primary sample.

\begin{figure}
\begin{centering}
\includegraphics[width=8.5cm]{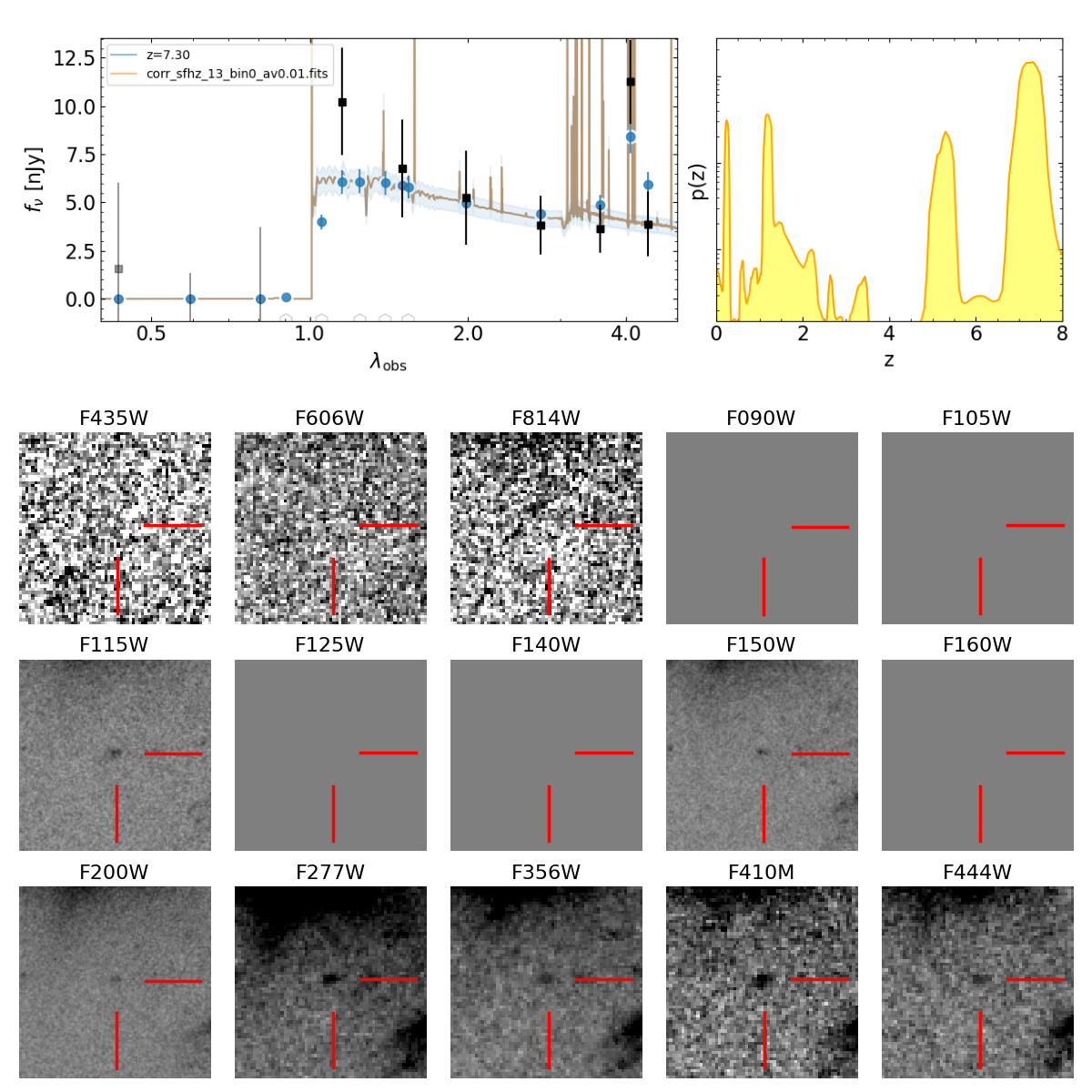}
\caption{Example of a $z\sim7$ [OIII] candidate not contained within the UNCOVER DR1 F277W+F356W+F444W selected catalog. Top left panel:  Best-fit \textsc{EA$z$Y} SED. Black points show measured fluxes and blue points show the corresponding best-fit \textsc{EA$z$Y} SED fluxes. 
 Top right panel: Photometric \textsc{EA$z$Y} redshift probability curve in arbitrary units, on a logarithmic scale.  Bottom panels: Postage stamps in each filter, $2.5$ arcseconds on a side. Blank stamps (F090W, F105W, F125W, F140W, F160W) indicate no bandpass coverage for this $z\sim7$ [OIII] candidate.}\label{missed_ex}
\par\end{centering}
\end{figure}

\begin{figure}
\begin{centering}
\includegraphics[width=8.5cm]{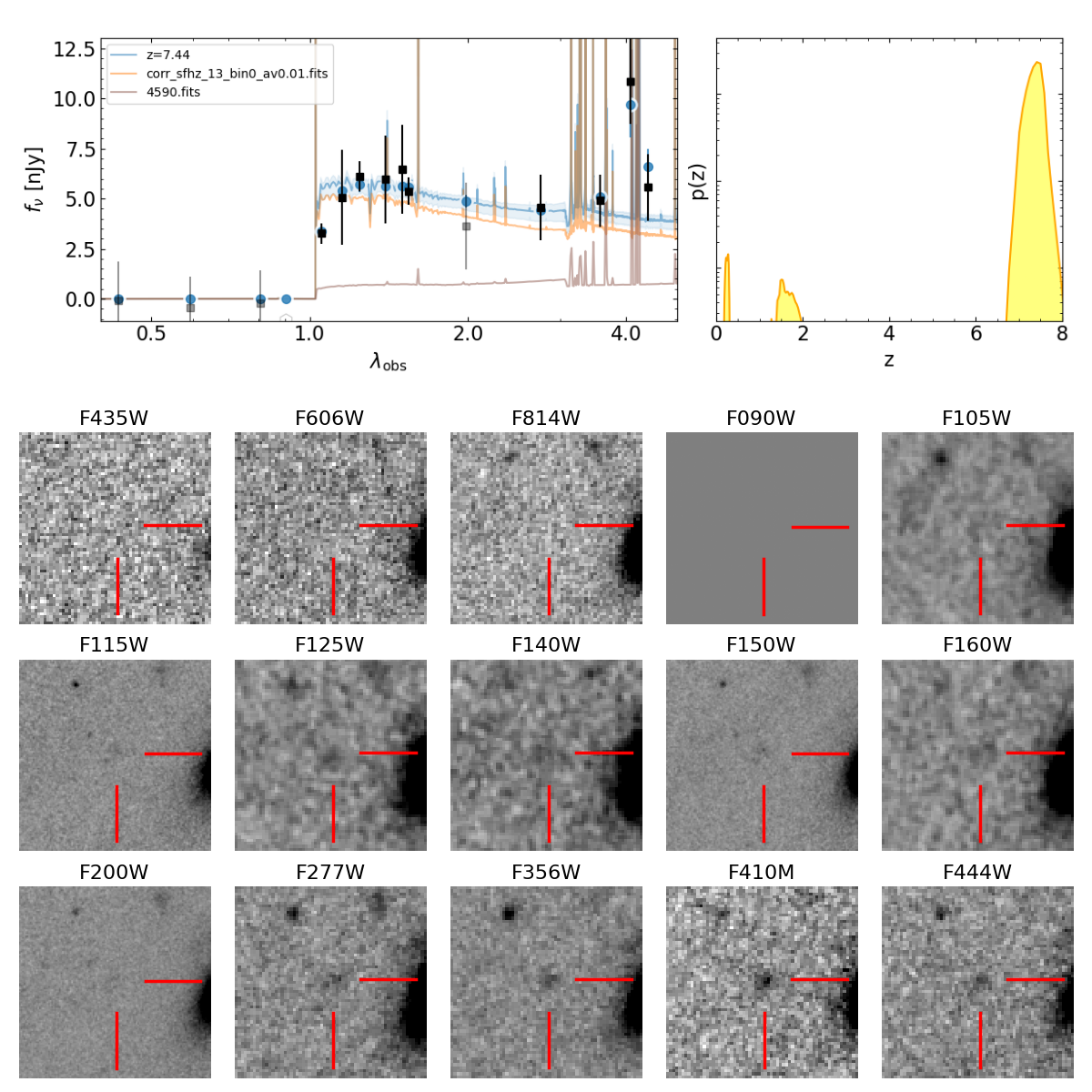}
\caption{Example of a $z\sim7$ [OIII] candidate flagged as having unreliable photometry due to SNR$_{\rm F444W}<3$ by the UNCOVER DR1 F277W+F356W+F444W selected catalog. See the caption of Figure \ref{missed_ex} for a description of the sub-panels. }\label{flagged_ex}
\par\end{centering}
\end{figure}

\section{comparison to the existing UNCOVER catalog}
Cross-matching to the F277W+F356W+F444W UNCOVER DR1 catalog, we find that $22\%$ ($N=15$) of our candidates have no cataloged counterpart.  Visually inspecting the $15$ sources without UNCOVER counterparts (see Figure \ref{missed_ex}), we find that many neighbor bright stars and galaxies, likely indicating that UNCOVER's removal of spurious sources also removed some real galaxies.   Only 39 out of our 68 candidates were designated by \citet{weaver24} as having reliable photometry (\texttt{USE\_PHOT}=1, see Figure \ref{flagged_ex}).  Objects can be flagged as unreliable for several reasons including bad-pixel artifacts, proximity to bright cluster galaxies, and low signal-to-noise ratio (SNR$_{\rm F444W}<3$).  The UNCOVER DR1 catalog includes flags to indicate which condition triggered the \texttt{USE\_PHOT}$=0$ flag, and we find that all our flagged candidates have low F444W SNRs (\texttt{FLAG\_LOWSNR}=1).  While spectroscopic confirmation is still needed, this suggests that searches for strong emission lines can find $z\sim7$ sources that would otherwise be missed even in extremely deep {\it JWST} broadband surveys.

\section{photometric redshifts}\label{photoz}

Photometric redshifts are not directly used in our [OIII] candidate selection.  We use photometric redshifts to justify limiting our [OIII] survey to the area with deep veto bandpass coverage.  We also compare our measurements to existing spectroscopic redshifts and the photometric redshifts in UNCOVER's DR1 catalog to help validate our color measurements. 

We measure photometric redshifts using a method similar to \citet{weaver24}.  We applied the \textsc{EA$z$Y} photometric software to all bands available for each object.  We used the \texttt{SFHZ\_CORR} SED templates which have $z$-dependent priors on allowable star-formation histories and a template based on the observed $z=8.5$ SMACS galaxy (ID=4590; e.g., \citet{rhoads23}).  We do not employ magnitude or $\beta$-slope priors.  We use a systematic flux error floor of $5\%$ and do not tune our photometric zeropoints to minimize photo-to-spec $z$ offsets. Our choice of \textsc{EA$z$Y} parameters attempts to select the most appropriate templates while avoiding fine-tuning which is not warranted for our $z\sim7$ emitters which have limited observational constraints.

\begin{figure}
\begin{centering}
\includegraphics[height=8.5cm]{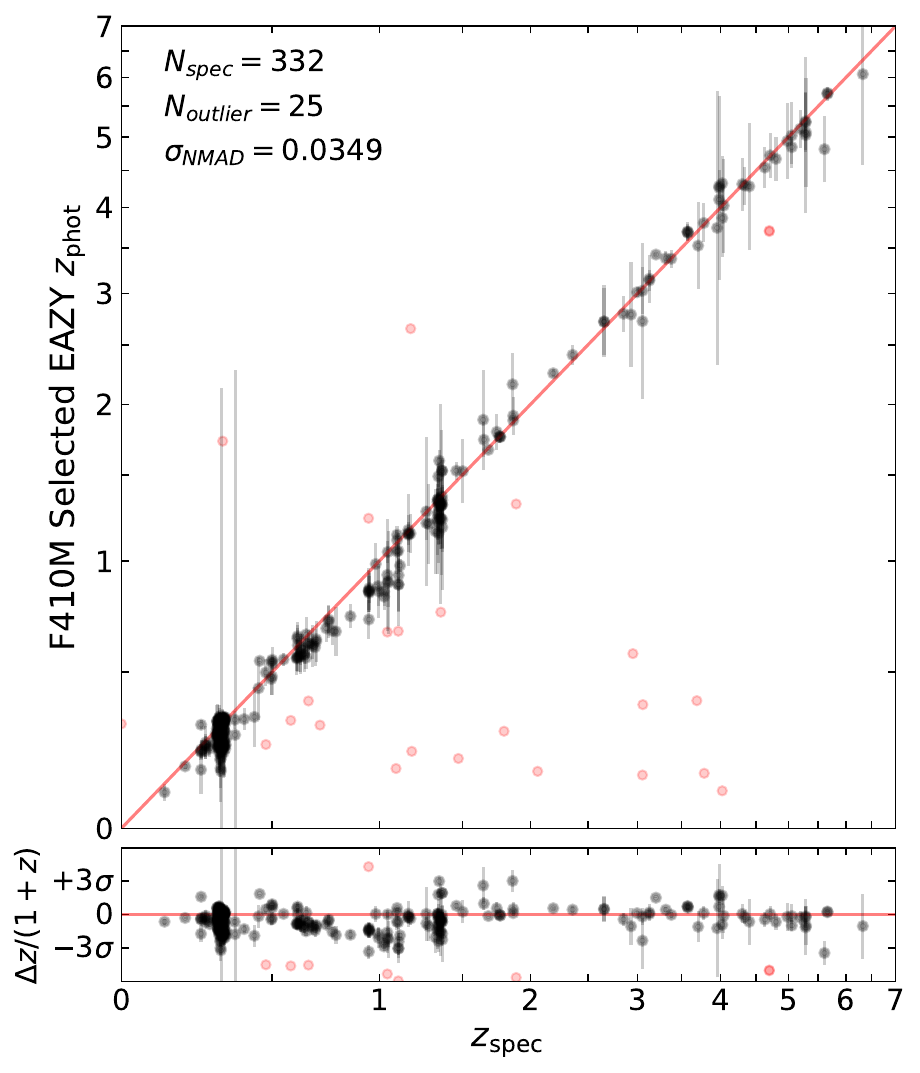}
\caption{Our F410M-selected \textsc{EA$z$Y} photometric redshifts compared to the UNCOVER spectroscopic sample. We find a $7.5\%$ outlier fraction (outlying sources shown by red points) and a $\Delta z$ spread of $\sigma_{\rm{NMAD}}=0.0349$.  The bottom panel shows the $(z_{\rm{phot}}-z_{\rm{spec}})/(1+z_{\rm{spec}})$ residuals normalized in units of $\sigma_{\rm{NMAD}}$.}\label{photoz_specz}
\par\end{centering}
\end{figure}

\begin{figure}
\begin{centering}
\includegraphics[height=8.5cm]{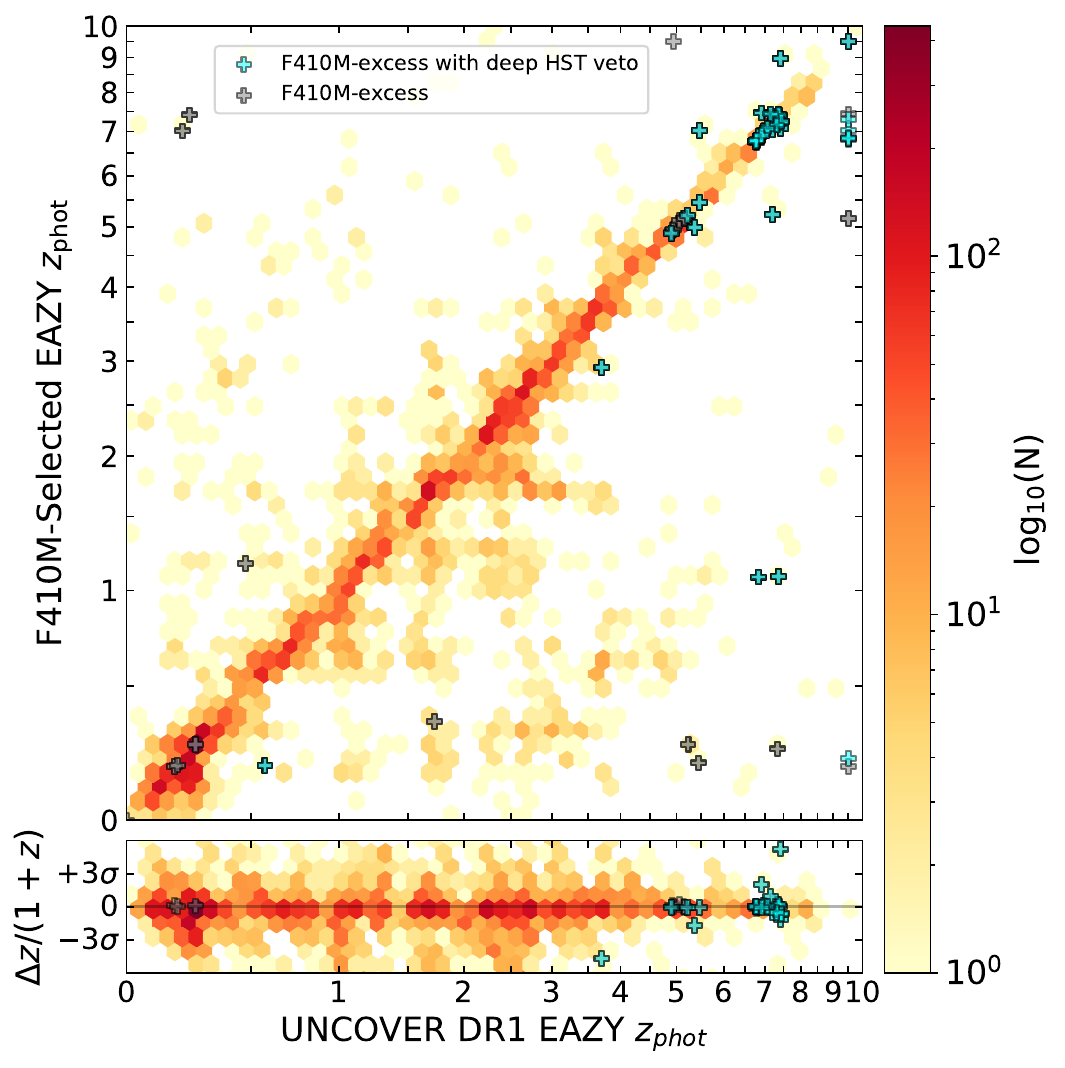}
\caption{Our photometric redshifts compared to UNCOVER DR1's published photometric redshifts.  [OIII] candidates not found by UNCOVER DR1's F277W+F356W+F444W selection are placed at $z=9.5$.  For our F410M-selected catalog, [OIII] candidates that have EA$z$Y photometric redshifts greater than $z=10$ are plotted at $z=9.5$. The bottom panel shows our $(z_{\rm{phot}}-z_{\rm{phot}}^{\rm{DR1}})/(1+z_{\rm{phot}}^{\rm{DR1}})$ residuals normalized in units of $\sigma_{\rm{NMAD}}$.
}\label{uncover_photoz}
\par\end{centering}
\end{figure}

In Figure \ref{photoz_specz}, we compare our computed photometric redshifts to the spectroscopic redshifts contained within the DR1 UNCOVER catalog \citep[compiled from many sources, with primary sources][]{treu15,richard21}.  For this comparison, we only consider F410M-selected objects with reliable photometry: SNR$_{\rm{F410M}}>4.5$ and outside of our bright star and edge mask.  We find a normalized median absolute deviation of $\sigma_{\rm{NMAD}}=0.0349$ and an outlier fraction of $7.5\%$ over a $N=332$ cross-matched sample within $r=0.3''$.  Here $\sigma_{\rm{NMAD}}$ is defined as:

\begin{equation}
\sigma_{\rm{NMAD}}=1.48\times \rm{median}\left(\frac{\vert \Delta z - \rm{median}(\Delta z)\vert}{1+z_{\rm{spec}}}\right)
\end{equation}

\noindent and the outlying fraction is defined as the fraction of objects with:

\begin{equation}
\vert z_{\rm{phot}} - z_{\rm{spec}}\vert \ge 0.015 (1+z_{\rm{spec}})
\end{equation}

\noindent Our results are comparable to the values found by \citet{weaver24}: $\sigma_{\rm{NMAD}}=0.0284$ and an outlier fraction of $7.5\%$.  

In Figure \ref{uncover_photoz}, we compare our \textsc{EA$z$Y} results to the UNCOVER DR1 cataloged photometric redshifts with reliable photometry (\texttt{USE\_PHOT}=1).  We find a normalized median absolute deviation of $\sigma_{\rm{NMAD}}=0.0352$ and an outlier fraction of $15.1\%$.  Unsurprisingly, the performance is somewhat degraded when considering the full photometric sample which -- unlike the $z_{\rm{spec}}$ sample -- is not preferentially drawn from relatively bright, nearby galaxies.  

Within Figure \ref{uncover_photoz}, we also highlight our $z\sim7$ [OIII] candidates.  As expected most candidates ($N=47$ out of $68$) have photometric redshifts at $z\sim7$ ($N=31$) or at $z\sim5$ ($N=16$).  If we restrict our sample to candidates that have deep veto band coverage ($N=33$; light blue symbols), we find that the number of $z_{\rm{phot}}\sim7$ candidates is $N=21$, while the number of $z_{\rm{phot}}\sim5$ candidates is only $N=5$.  Thus, we find that the deep veto restriction increases the selection percentage of $z_{\rm{phot}}\sim7$ candidates from $46\%$ to $64\%$, indicating the importance of these deep {\it HST}-ACS bands for isolating a clean sample of $z\sim7$ [OIII] emitters.

\begin{figure}
\begin{centering}
\includegraphics[width=8.5cm]{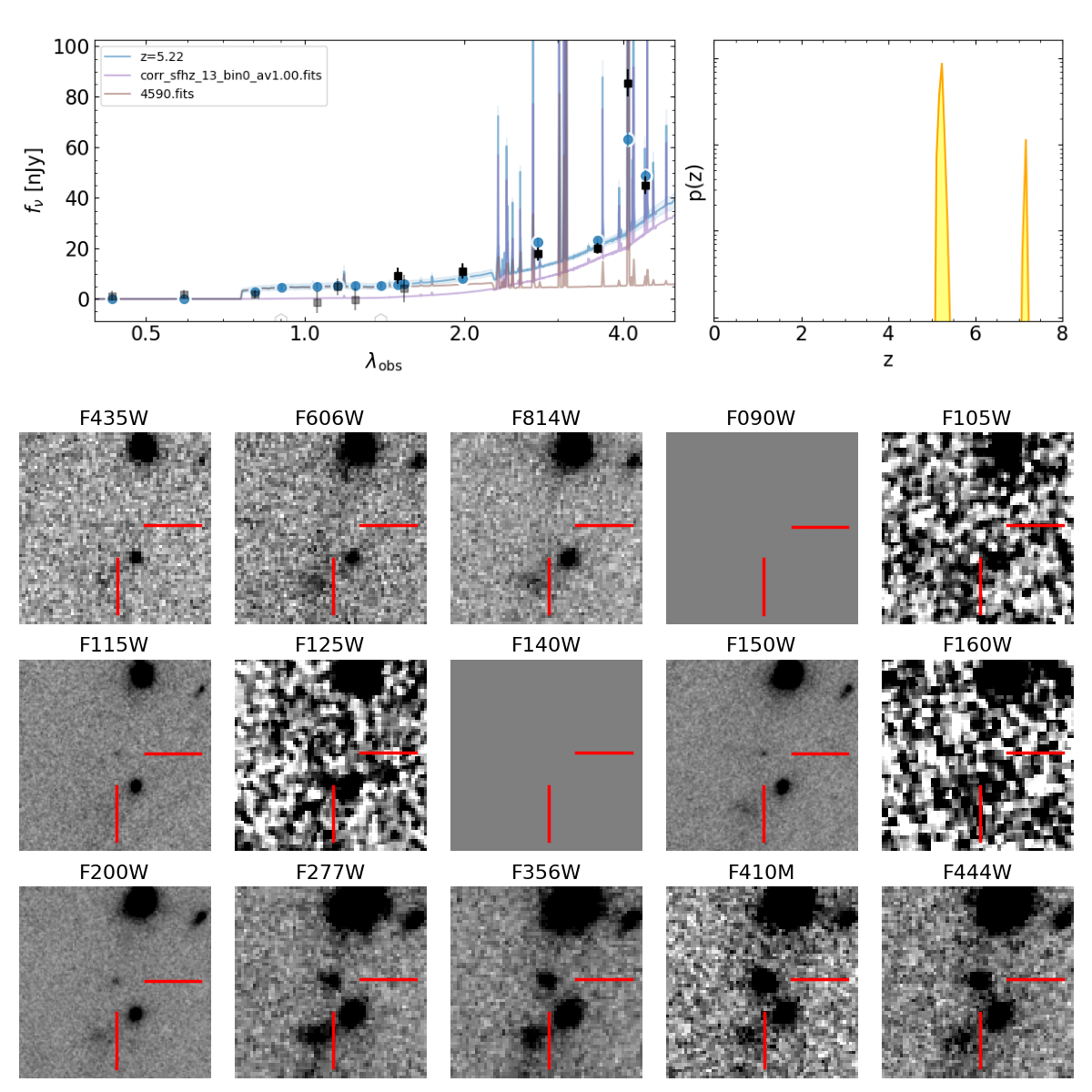}
\caption{Example of a $z\sim7$ [OIII] candidate with an EAzY photometric redshift of $z\sim5$. See the caption of Figure \ref{missed_ex} for a description of the sub-panels.}\label{ha_ex}
\par\end{centering}
\end{figure}

In Figure \ref{ha_ex}, we show an example of a selected $z\sim7$ [OIII] candidate that has $z_{\rm{phot}}\sim5$.  The \textsc{EA$z$Y}  probability distribution shows that there is a slightly disfavored alternative fit at $z\sim7$.  This is the general case for candidates that have $z_{\rm{phot}}\sim5$ with deep {\it HST} veto coverage.   When deep veto bandpasses are not available, we find a handful of candidates that have photometric solutions that highly disfavor $z_{\rm{phot}}\sim7$.  This is caused by $z_{\rm{phot}}\sim5$ sources with very bright F277W counterparts that are best-fit by [OIII] falling within this bandpass.  These photometric fits again emphasize the need for deep veto bandpasses, and we minimize foreground contamination by limiting our analysis to our primary sample of 33 objects with deep {\it HST}-ACS coverage.  Using the three SED figures to help illustrate this restriction, the [OIII] candidate in Figure \ref{missed_ex} lacks deep {\it HST}-ACS coverage and is excluded, while the [OIII] candidates in Figure \ref{flagged_ex} and \ref{ha_ex} have deep veto coverage (F814W $5\sigma$ depth $>28$ AB) and are included in our primary sample.

\section{$z\sim7$ [OIII] Size Distribution and [OIII] Pairs}

\begin{figure}
\begin{centering}
\includegraphics[width=8.5cm]{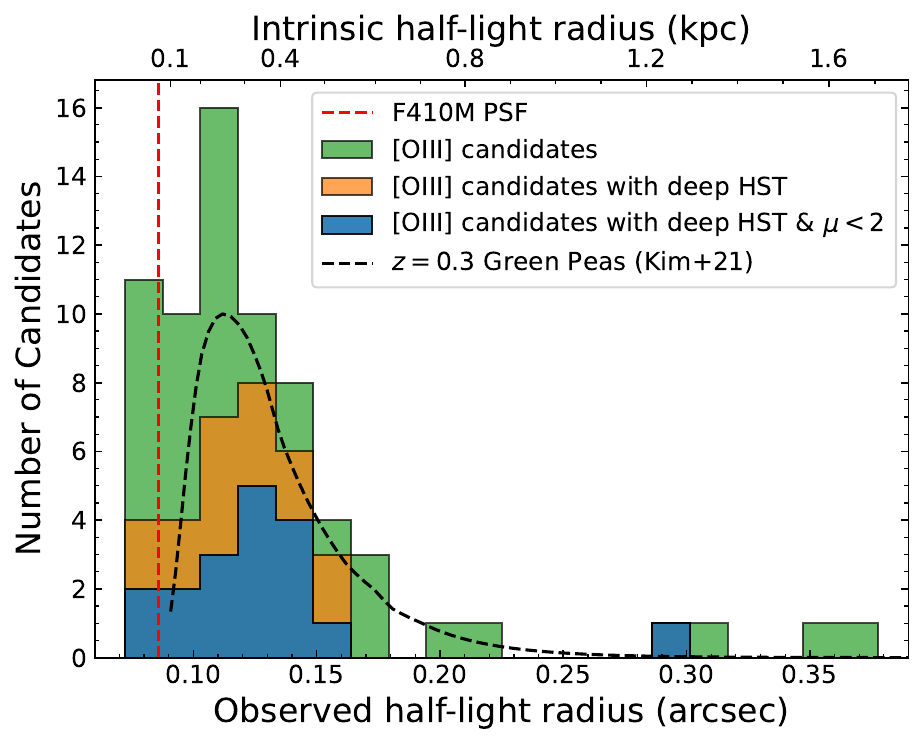}
\caption{The observed F410M size distribution of all $z\sim7$ [OIII] emitter candidates (green histogram).  We indicate the subset of candidates with deep HST data used to compute the luminosity function (orange histogram).  We also show this same distribution with highly magnified sources ($\mu>2$) removed (blue histogram). 
 In the order listed, the median observed half-light radii are $0.12''$, $0.13''$, and $0.13''$. For comparison, we show the best-fit log-normal intrinsic size distribution of $z\sim0.3$ Green Peas from \citet{kim21}.  We have normalized this distribution to a peak value of 10 for display purposes. }\label{o3_size}
\par\end{centering}
\end{figure}

The typical [OIII] candidate is very compact with a median observed F410M half-light radius of $\sim0.13''$.  Figure \ref{o3_size} shows the size distribution of the $z\sim7$ [OIII] emitter candidates. Gravitational lensing could bias this distribution by making our sample appear more extended. Lensing conserves an object's surface brightness by increasing both its area and flux by the magnification factor, $\mu$.  We remove highly magnified sources ($\mu>2$ as identified in the \citealt{furtak23} lensing maps) and find that the median observed F410M half-light radius is unchanged, suggesting that magnification is not significantly biasing our median size measurement.  

We estimate the intrinsic size of our candidates by convolving a S\'ersic exponential disc with the F410M PSF.  We find an observed F410M half-light radius of $0.13''$ indicates an intrinsic half-light radius of $r_{1/2} = 0.4$ kpc at $z=7$.  This compact size is similar to local $z\sim0.3$ strong [OIII] emitters that have a typical radius of 0.33 kpc \citep{kim21}, potentially indicating little size evolution for EW([OIII])$\gtrsim500$ \AA\ sources. However, \citeauthor{kim21} measure the rest-frame UV size, while we measure the rest-frame optical size within a medium bandpass expected to be dominated by [OIII] emission.  Additionally, we make a simplifying assumption that our sources are S\'ersic exponential discs, rather than fitting the S\'ersic index as a free parameter.  We defer further morphology analysis to a future paper but conclude that the typical [OIII] candidate is very compact, and we use the median size in our completeness simulations described in Section \ref{scomp}.

Within our sample of $N=68$ [OIII] candidates, three systems are pairs, and one system is a triplet.  Candidates are identified as pairs or triplets by having a neighboring [OIII] candidate within $2''$ or $\sim10$ kpc assuming the sources are at the same $z=7$ redshift.  All systems with potential close companions lack deep HST-ACS data, giving us less confidence in our ability to distinguish between $z=5$ H$\alpha$ emitters and $z=7$ [OIII] emitters.  As previously discussed in Section \ref{photoz}, we mitigate H$\alpha$ contamination by limiting our primary sample to the $N=33$ candidates with deep HST-ACS coverage.  This results in the exclusion of these pair/triplet systems from our number counts and luminosity function analysis.

\section{$z\sim7$ [OIII] Number Counts}
\begin{figure}
\begin{centering}
\includegraphics[width=8.5cm]{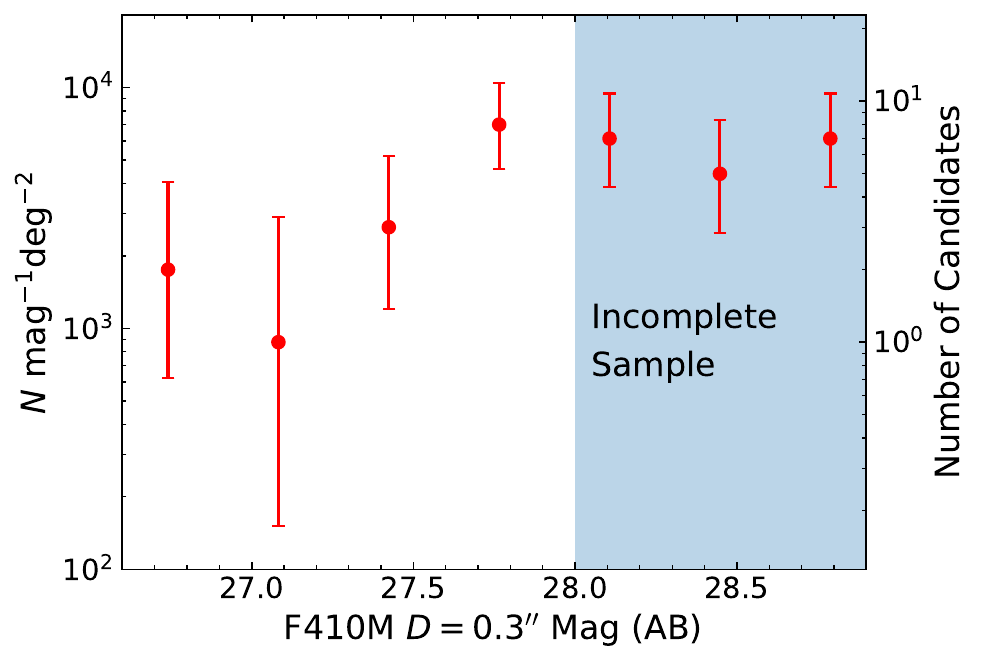}
\caption{Number counts of $z\sim7$ [OIII] emitter candidates.  Error bars show the $1\sigma$ Poisson uncertainty. The y-axis on the right shows the number of $z\sim7$ [OIII] candidates in each $0.3$ mag bin. Our sample begins to become incomplete at F410M $\gtrsim 28$ mag (blue-shaded region).
}\label{counts}
\par\end{centering}
\end{figure}

As a prelude to computing the $z\sim7$ [OIII] luminosity function, in Figure \ref{counts}, we display their number counts as a function of F410M magnitude.  We only consider the survey area with deep {\it HST} veto band coverage ($13.68$ arcmin$^2$), and we do not account for gravitational lensing which is found to magnify most sources within the survey area by a factor of $\gtrsim2$ \citep{weaver24}. Our candidate selection includes a $\Sigma>2$ cut (see Equation \ref{sigeq}) which begins to limit our sample at F410M $\gtrsim 28$ mag (see Figure \ref{uncover_select}).  At this magnitude and higher we expect our sample to be incomplete.  This effect is not accounted for in our computed number counts.  In the next section, we present completeness simulations that account for gravitational lensing and our selection criteria, allowing us to compute the [OIII] luminosity function.

\section{Completeness Simulations}\label{scomp}

We insert simulated [OIII] emitters into the F410M and F444W science images and then extract, measure, and select [OIII] candidates using our standard procedure described in Section \ref{select}.   We inserted $100$ simulated [OIII] emitters per run and performed $500$ independent runs, giving us a total of $50,000$ simulated sources. [OIII] emitters are inserted into the survey area with deep {\it HST} veto band coverage that is not excluded by our bright star mask ($13.68$ arcmin$^2$).  This area is also used when computing our LF.

Simulated emitters are characterized by three parameters: [OIII] luminosity, EW, and redshift.  We randomly populate these parameters from realistic distributions to capture potential observational biases within our completeness corrections. 

We randomly select [OIII] line luminosities from a power-law distribution $dN/dL_{[OIII]}\propto L_{[OIII]}^{-2.0}$.  This power law is consistent with the $z\sim6$ UV faint-end slope of $\alpha=-2.0$ \citep[e.g.,][]{bouwens15,bouwens21,finkelstein15} and is used by \citet{matthee23} in their $z\sim6$ [OIII] LF fit.  Our power-law distribution does not enforce an exponential cutoff at $L^{*}$ giving us better recovery-fraction measurements at high luminosities.  The minimum observed line luminosity was set to be below the field's depth at $10^{41.5}$ erg s$^{-1}$.  These observed luminosities are modified by a factor of $1/\mu$ to obtain the intrinsic values, where the magnification factor, $\mu$, is obtained from the \citet{furtak23} lensing maps.

We selected [OIII] rest-frame EWs by randomly sampling a declining exponential with a scale length $W_{0}=1000$\AA\ and a minimum EW threshold of $300$\AA: $dN/dEW_{[OIII]}\propto e^{-EW/W_{0}}$.  The distribution's scale length is in line with expectations from archival IRAC and {\it JWST} studies \citep{Labbe13,smit15,endsley21,matthee23}.

To model filter profile effects, we uniformly distribute sources in comoving volume over the redshift range that the F410M filter covers, $z=6.5$ to $7.8$ \citep[for a similar procedure see][]{konno18,taylor20,taylor21,ning22}.  The source plane's coordinates were uniformly sampled to determine the simulated object's location.  Using the \citet{furtak23} deflection and magnification maps, we projected the objects onto the image plane.  This was done for each pixel within the morphology postage stamp using nearest-neighbor mapping.

To facilitate our nearest-neighbor mapping, we generated high-resolution morphology stamps with a $0.013''$ pixel scale ($3\times$ the native pixel scale of $0.04''$). Simulated emitters have an exponential disc morphology with an intrinsic half-light radius of $r_{1/2}=0.4$ kpc.  This half-light radius was set to match the observed median size of our $z\sim7$ [OIII] candidates ($r_{1/2}=0.13''$) once this disc profile is convolved with the F410M PSF.  Over the extent of the F410M redshift coverage, the physical scale varies from $5.4$ - $5.0$ kpc per arcsec. We maintain a relatively constant half-light radius by producing a library of $r_{1/2}=0.4$ kpc disc profile stamps at redshifts $z=6.7$, $7.0$, $7.3$, and $7.6$ selecting the closest stamp in redshift proximity for each simulated emitter's randomly selected redshift.

Given our sampled values of redshift, line luminosity, and EW, we compute the F410M and F444W flux by using the known filter profiles and our assumed [OIII] emitter spectral shape (see Section \ref{ewselect} for details).  For each simulated object, we 1) scale our [OIII] morphology to the desired flux, 2) map it to the image plane, 3) convolve the morphology with the F410M or F444W PSF, and 4) insert it into our F410M and F444W science images.  While bluer bandpasses have no sources inserted, we still perform aperture photometry on these bandpasses in the same manner as described in Section \ref{select}.  This practice ensures that our completeness corrections account for the survey area blocked by foreground sources, and therefore we do not expect to recover all inserted emitters even at the brightest luminosities.     

\begin{figure}
\begin{centering}
\includegraphics[width=8.5cm]{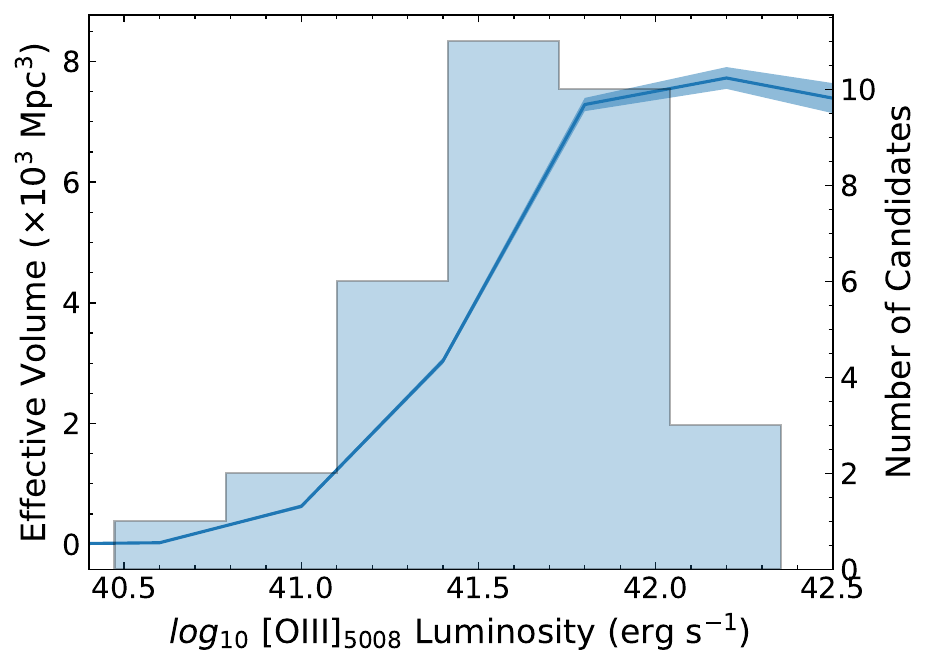}
\caption{The effective survey volume as a function of [OIII] luminosity. The shaded region around the curve represents the $\pm1\sigma$ Poisson uncertainties of our completeness simulations. The distribution of $z\sim7$ candidates (blue histogram and right y-axis) is shown for comparison.}\label{veff}
\par\end{centering}
\end{figure}

Our candidate selection is described by Equation \ref{seleq}.  Rather than applying an EA$z$Y $z_{\rm{photo}}$ cut,  we mitigate low-redshift interlopers by limiting the LF survey area to regions with deep {\it HST}-ACS coverage (as discussed in Section \ref{photoz}). 
Our recovered [OIII] luminosities are computed by assuming a nominal redshift of $z=7.17$.  The luminosity error that this convention can introduce is small (0.1 dex) relative to our luminosity bins (0.4 dex) used to construct the LF.  Additionally, our completeness simulations attempt to account for the input-to-recovered luminosity scatter by adopting realistic input distributions and recording the ratio of the number of recovered [OIII] emitters measured to fall within a given luminosity bin versus the number of input [OIII] emitters known to fall within the same luminosity bin.  In other words, the loss of redshift information is modeled with our completeness simulations.  

Given this nominal redshift, the F410M/F444W filter responses, and an [OIII] emitter spectral shape (see Section \ref{ewselect}), we compute the needed [OIII] luminosities to match the measured F410M/F444W total magnitudes. For each object, the PSF-matched $0.3''$ diameter aperture F410M and F444W flux is converted to a total flux by applying a $f^{\rm{F410M}}_{\rm{AUTO}}/f^{\rm{F410M}}_{D=0.3''}$ correction factor.  Our completeness simulation employs the same method for measuring [OIII] luminosities as used for our main sample consisting of $N=33$ [OIII] candidates.

\section{Luminosity function}

We compute the effective survey volume, $V_{\rm{eff}}$, as a function of [OIII] luminosity via:

\begin{equation}
V_{\rm{eff}}=\int_{0}^{\infty}\int_{\mu>\mu_{min}}\frac{dV_{\rm{com}}}{dz}f(z,\mu,L_{[OIII]})d\Omega(z,\mu)dz
\end{equation}

\noindent where $V_{\rm{com}}$ is the comoving volume, $f$ is the recovery fraction - the number of recovered versus input [OIII] emitters determined by our completeness simulations, and $d\Omega$ is the solid angle element at a given redshift and magnification where $\mu_{\rm{min}}$ is the minimum magnification at which an emitter with [OIII] luminosity $L_{[OIII]}$ can be successfully recovered.  For a similar calculation of $V_{\rm{eff}}$ for lensing fields see \citet{atek15,atek18}.

For a simpler candidate selection, we do not fold in photometric redshifts and assume a redshift of $z=7.17$ (corresponding to the F410M pivot wavelength) for [OIII] luminosity measurements. To account for this assumption within our completeness simulations, our recovery fractions record the number of selected objects measured to fall within a given luminosity bin (where $z=7.17$ is used) relative to the known number of input objects at that same luminosity bin (where $z_{\rm{input}}$ is used).  The $V_{\rm{eff}}$ integral was evaluated over a redshift range of $6.72<z<7.59$ which corresponds to the half-power wavelengths of the F410M passband. Consistent with our number counts calculation, we only consider the
survey area with deep {\it HST} veto band coverage (13.68 arcmin$^2$) to minimize foreground contamination concerns.

In Figure \ref{veff}, we show that the effective survey volume peaks at $8\times10^{3}$ Mpc$^{3}$ and quickly declines for emitters with luminosities $L_{[OIII]}<10^{41.5}$ erg s$^{-1}$.  In addition, this figure shows that the distribution of $z\sim7$ [OIII] candidates also follows this $L_{[OIII]}$ trend as expected.    

To compute the $z\sim7$ [OIII] luminosity function, we interpolate to find $V_{\rm{eff}}(L_{\rm{[OIII]}})$ for each candidate emitter and use the binned differential luminosity function estimator

\begin{equation}
\phi(\left\langle L_{{\rm {[OIII]}}}\right\rangle )=\frac{1}{\Delta L_{{\rm {[OIII]}}}}\sum_{k}\frac{1}{V_{{\rm {eff}}}(L_{{\rm {[OIII]}}})}
\end{equation}

\noindent where $\left\langle L_{{\rm {[OIII]}}}\right\rangle$ is the average [OIII] luminosity of a bin, $\Delta L_{\rm {[OIII]}}$ is the bin width, and the summation is over all sources $k$ within the luminosity bin. When computing the luminosity function, we only include emitters with a completeness of $>5\%$ to mitigate completeness correction uncertainties.  This removes two sources resulting in a final luminosity function sample size of 31 [OIII] candidates.  

\begin{figure}
\begin{centering}
\includegraphics[width=8.5cm]{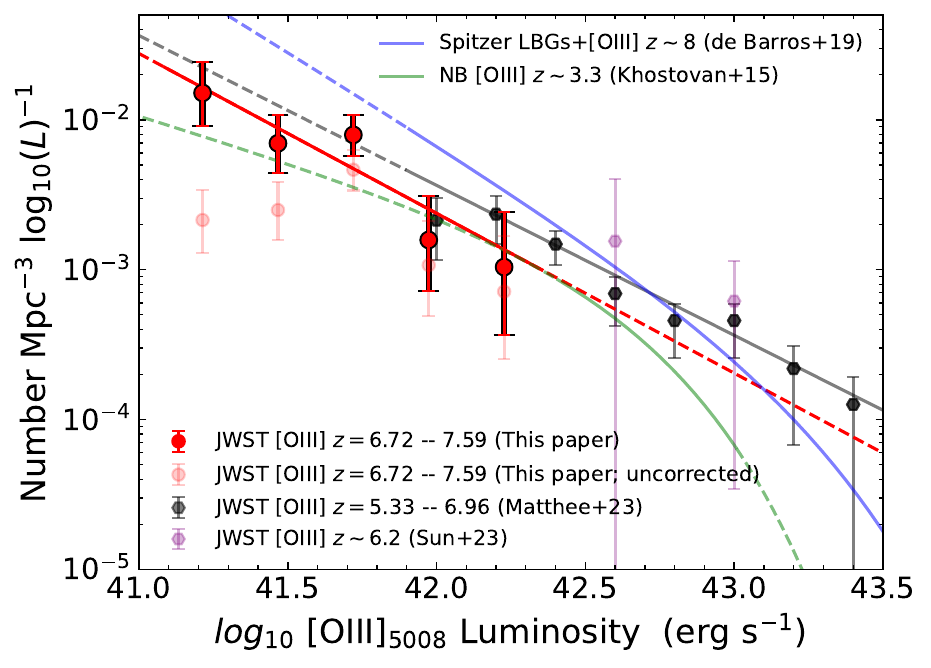}
\caption{Our $z\sim7$ [OIII] luminosity function with $\pm1\sigma$ Poisson error bars in comparison to archival studies. We fit our LF with a power law (red line), finding a faint-end slope of $\alpha=-2.07^{+0.22}_{-0.23}$. Best-fit LFs are shown as solid curves over the luminosity range constrained by data points.  Remarkably little evolution is seen in the [OIII] LF from $z\sim3$ to $\sim8$. }\label{lf}
\par\end{centering}
\end{figure}

In Figure \ref{lf}, we show our $z\sim7$ results in comparison to the $z\sim6$ [OIII] luminosity function from \citet{matthee23}.  We find that our results are consistent over the luminosity range of overlap $41.9>\rm{log}_{10}L_{\rm{[OIII]}}>42.4$ tentatively suggesting little evolution from $z=6$ to $7$ a $165$ Myr duration.  
The survey volume is an order of magnitude less than the $z\sim6$ survey ($10^{5}$ Mpc$^3$), explaining the lack of relatively rare $L_{\rm{[OIII]}}>10^{42.5}$ erg s$^{-1}$ sources within the $z\sim7$ survey.  We note that the $z\sim6$ LF is derived from a field that contains a luminous quasar at $z=6.33$ and likely probes a peak in the cosmic number density. While \citeauthor{matthee23} mitigate this effect by masking $z=6.3 - 6.35$, a $\sim20$ Mpc extent from their survey, we know that high-redshift proto-clusters can extend out to $\sim60$ Mpc  \citep{hu21}.  Thus, there remains some concern that their $z\sim6$ number densities could be enhanced. 

Using Cash Statistics \citep{cash79}, we fit our luminosity function with a power law:

\begin{equation}
\Phi(L)dL=\left(\frac{L}{10^{42}}\right)^{\alpha}10^{\beta}dL,
\end{equation}

\noindent finding $\alpha=-2.07^{+0.22}_{-0.23}$ and $\beta=-45.0^{+0.1}_{-0.1}$ as shown by a red line in Figure \ref{lf}. We compute $1\sigma$ errors with a Monte Carlo simulation that perturbs our LF data by Poisson random deviates.  This procedure is used to create $N=10000$ perturbed LFs.  We perform our standard LF fitting and find the inter 68th percentile of the resulting distributions of $\alpha$ and $\beta$.

We find no dramatic decline or break in our $z\sim7$ luminosity function which covers a luminosity range of $41.1<\rm{log}_{10}(L/\rm{erg\,s}^{-1})<42.4$. For this reason, we did not fit a Schechter function \citep{schechter76}:
\begin{equation}
\Phi(L)dL=\phi^{*}\left(\frac{L}{L^{*}}\right)^{\alpha}e^{-L/L^{*}}d\left(\frac{L}{L^{*}}\right).
\end{equation}

\noindent which has a third free-parameter, $L^*$, that defines the luminosity above which the distribution departs from a power law by declining exponentially.   

Comparing our LF fit to \citeauthor{matthee23}'s best fit Schechter function at $z\sim6$ 
($\alpha,\rm{log}_{10}\phi^*,\rm{log}_{10}L^*=-0.24^{+1.24}_{-1.88},-2.77^{+0.17}_{-5.27},42.18^{+3.49}_{-0.33}$), we find that our LF fit does not suggest a shallow faint-end slope of $\alpha=-0.24$ as found by their best-fit value.  A shallow faint-end slope at high redshifts could be explained by the mass-metallicity relation causing lower mass galaxies to become oxygen-deficient ($12+$log$_{10}$(O/H)$\lesssim 7.5$; e.g., \citealt[][]{kojima20}), resulting in fewer emitted [OIII] photons. Instead, the UNCOVER survey - which is an order of magnitude deeper - constrains the faint-end slope to be within a $\pm1\sigma$ range of $-1.85$ to $-2.30$ with no evidence for a turnover.

\begin{figure}
\begin{centering}
\includegraphics[width=8.5cm]{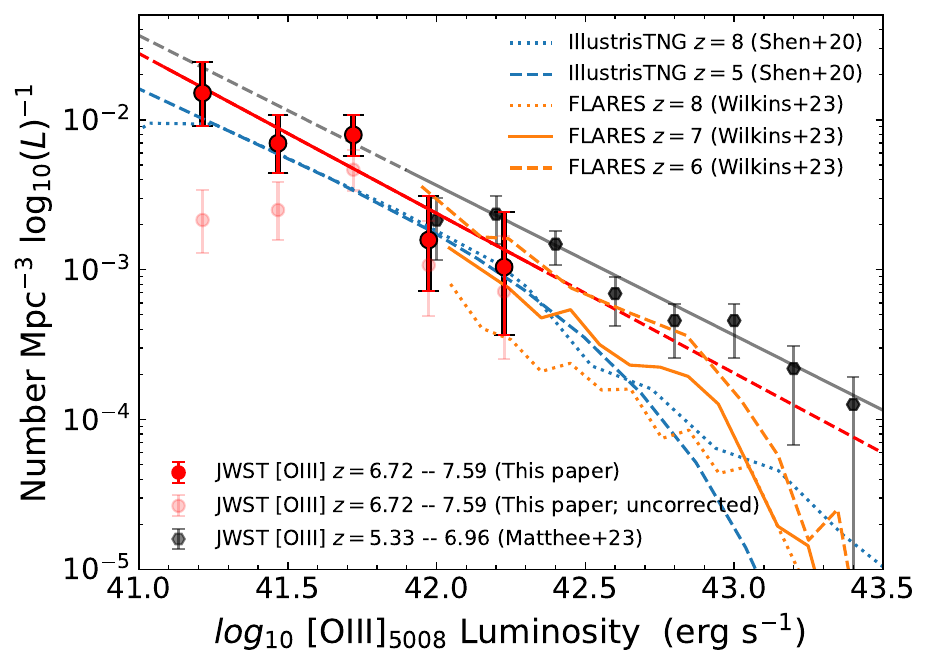}
\caption{The $z\sim6$ and $z\sim7$ [OIII] luminosity functions in comparison to simulations. Simulations favor little to no evolution from $z=6$ to $7$, consistent with our results.  The IllustrisTNG results span our [OIII] luminosity range and predict a steep faint-end slope which is also consistent with our $z\sim7$ [OIII] luminosity function.  The IllustrisTNG simulations predict [OIII]+H$\beta$ LF results. Consistent with our LF computations, we convert these to [OIII]$_{5008}$ LFs by assuming that the 5008 line contributes $64\%$ of the total [OIII]+H$\beta$ line flux.  }\label{lfsim}
\par\end{centering}
\end{figure}

\citeauthor{matthee23} also explore an alternative fit, where the faint-end slope is fixed to the $z\sim6$ UV LF slope of $-2$ \citep[e.g.,][]{bouwens15,bouwens21,finkelstein15}.  This assumption facilitates comparison to our results because our best-fit slope, $-2.07^{+0.22}_{-0.23}$, is $1\sigma$ consistent with this value.  With a fixed $-2$ slope, \citeauthor{matthee23} find best-fit Schechter parameters of log$_{10}\phi^*=-7.74^{+4.01}_{-0.14}$ and log$_{10}L^*=46.94^{+0.05}_{-3.90}$.    We note that their best-fit $L^*$ value favors an exponential decline at luminosities beyond their surveyed range ($41.9<\rm{log}_{10}(L/\rm{erg\,s}^{-1})<43.5$), suggesting that a larger survey volume is needed to place meaningful constraints on $L^*$ which is a parameter highly anti-correlated with fitted $\phi^*$ values.  Given their uncertain best-fit $L^*$ value, we independently fit \citeauthor{matthee23}'s $z\sim6$ [OIII] LF with a $-2$ fixed slope power law and find $\beta = -44.9\pm0.1$. Using the same $-2$ fixed slope and our $z\sim7$ LF, we find a best fit $\beta=-45.0\pm0.1$. This is consistent with a $\sim1.25$ factor decline from $z\sim6$ to $7$ in the luminosity density, $\rho^{\text{[OIII]}}$:

\begin{equation}
\rho_{\text{{[OIII]}}}=\int^{L_{\rm{max}}}_{L_{\rm{min}}} L\Phi(L)dL,
\end{equation}

\noindent where we have integrated over the same luminosity range for both the $z\sim6$ and $z\sim7$ power-law luminosity function, and 

\begin{equation}
\frac{\rho^{z\sim6}_{\text{{[OIII]}}}}{\rho^{z\sim7}_{\text{{[OIII]}}}}=10^{\Delta\beta}\sim1.25,
\end{equation}

\noindent This small $\sim1.25$ decline of $\rho_{\text{[OIII]}}$  is within the $1\sigma$ error for this parameter, showing no evidence for significant luminosity density evolution over this $165$ Myr duration in elapsed cosmic time. We note that a small $\sim1.25$ decline in luminosity density is seen in the UV LF from $z=6$ to $7$ \citep[e.g.,][]{bouwens22}, potentially hinting at a similar evolution over this redshift range. However, a larger overlapping luminosity range than $41.9>\rm{log}_{10}L_{\rm{[OIII]}}>42.4$ is needed to robustly constrain the behavior of $\rho_{\text{[OIII]}}$ from $z=6$ to $7$.  Using our observed luminosity bounds, $41.1<\rm{log}_{10}(L/\rm{erg\,s}^{-1})<42.4$, and our best-fit power-law parameters, $\alpha=-2.07^{+0.22}_{-0.23}$ and $\beta=-45.0^{+0.1}_{-0.1}$, we find an observed luminosity density of log$_{10}\rho_{\text{[OIII]}}=39.5\pm0.1$ erg s$^{-1}$ Mpc$^{-3}$ at $z\sim7$.

In Figure \ref{lf}, we also compare our results to the [OIII] LFs at $z\sim3.3$, $\sim6.2$, and $\sim8$ \citep{khostovan15,sun23,debarros19}.  To compare the H$\beta+$[OIII] LFs with the [OIII]$_{5008}$ LFs, we have assumed that the [OIII]$_{5008}$ line contributes $64\%$ of the total flux.  This is consistent with the literature \citep{sun23,matthee23} and with our [OIII]$_{5008}$ luminosity computation (see Section \ref{ewselect}).

While our results have no luminosity overlap with the $z\sim3$ survey, comparison between the $z=3$ and $z=6$ [OIII] LF shows very little evolution despite the UV luminosity density declining by a factor of $\sim4$  over the same redshift range \citep{finkelstein15,bouwens15,bouwens22}.   Additionally, the H$\beta+$[OIII] emitters at $z=3$ are selected to have EWs greater than $25$\AA\ while the typical $z\gtrsim6$ emitter has a much higher EW of $\gtrsim500$\AA.  Matching the $z=3$ sample to this same EW threshold would significantly reduce their number density and require a substantial luminosity density increase from $z=3$ to $6$. This further emphasizes that the evolution of strong [OIII] emitters is behaving differently than UV-selected galaxies over the $3<z<6$ redshift range. This may be explained by high-redshift star-forming galaxies producing more ionizing flux and higher [OIII] line luminosities at fixed SFR \citep[for further discussion,][]{sun23, matthee23}.

Compared to our results, the $z\sim8$ [OIII] LF \citep{debarros19} suggests higher number densities (by a factor of $\sim2$) at intermediate luminosities of $10^{42}$ erg s$^{-1}$ but has number densities roughly in agreement with the $z\sim6$ LF at $>10^{43}$ erg s$^{-1}$.  Overall, we find that our $z\sim7$ LF is in close agreement with the $z\sim6$ LF over the small range of overlapping luminosity, and we use our more sensitive F410M-excess selection to place the best constraints on the faint-end, finding a slope of $-2.07^{+0.22}_{-0.23}$.

In Figure \ref{lfsim}, we compare the $z\sim6$ and $z\sim7$ [OIII] LF to the IllustrisTNG \citep{shen20} and the FLARES \citep{wilkins23} simulations.  Studies of the bright-end of the $z\sim6$ [OIII] LF have found a factor of $\sim10$ higher observed number densities compared to simulations \citep{sun23}.  Our faint-end LF results are in closer agreement with simulations.  Computing the IllustrisTNG [OIII] number density at $z=8$ over our survey's luminosity range, we find log$_{10}n_{\text{[OIII]}}=-2.2$ Mpc$^{-3}$ which is $\pm1\sigma$ consistent with our number density result of log$_{10}n_{\text{[OIII]}}=-2.1\pm0.1$  Mpc$^{-3}$.  Additionally, we find a faint-end slope of $\alpha \sim -2$ which is roughly consistent with IllustrisTNG. The IllustrisTNG and FLARES simulations also predict little or no LF evolution from $z\sim6$ to $z\sim7$, which is consistent with our results.  However, as mentioned earlier, a larger $z\sim7$ survey volume is needed to more accurately constrain the bright-end of the LF and its overall evolution.

\begin{figure}
\begin{centering}
\includegraphics[width=8.5cm]{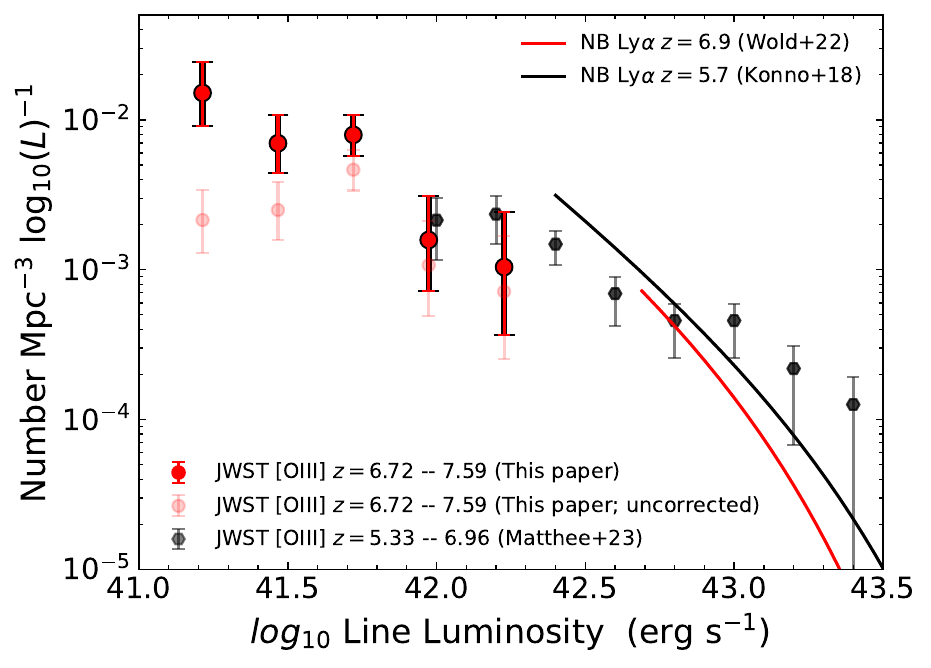}
\caption{Comparison of the [OIII] LFs to the Ly$\alpha$ LFs at $z\sim6$ and $z\sim7$.  See Section \ref{slya} for discussion.}\label{lya}
\par\end{centering}
\end{figure}

\begin{figure*}
\begin{centering}
\includegraphics[height=6cm]{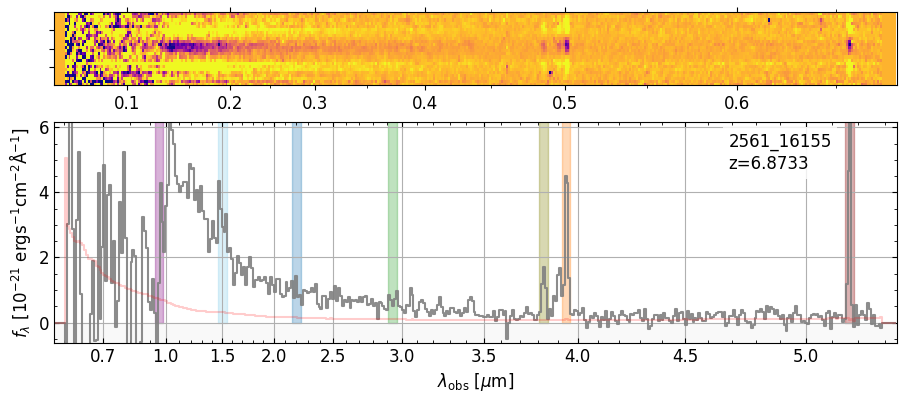}
\caption{\textit{JWST} NIRSpec/PRISM confirmation of a $z\sim7$ [OIII] candidate ($\alpha=3.58296$, $\delta=-30.39523$, $\mu=11.5$, F410M $=21.9$ AB, F444W $=23.4$ AB).  The top panel shows the 2-D spectra with the rest-frame wavelength shown on the x-axis.  The bottom panel shows the corresponding 1-D spectrum with UNCOVER program ID (2561), UNCOVER object ID (16155), and redshift ($z=6.8733$) displayed.  From left to right, we highlight the wavelength locations of Ly$\alpha$, CIII]$_{1907, 1909}$, MgII$_{2796, 2803}$, [OII]$_{3727, 3730}$, H$\beta$, [OIII]$_{5008}$, and H$\alpha$.  The light red curve shows the $1\sigma$ error array.  The format of this figure is modified from \textsc{msaexp} which is a software module for extracting and displaying {\it JWST} NIRSpec MSA spectra.      }\label{spec1}
\par\end{centering}
\end{figure*}
\begin{figure*}
\begin{centering}
\includegraphics[height=6cm]{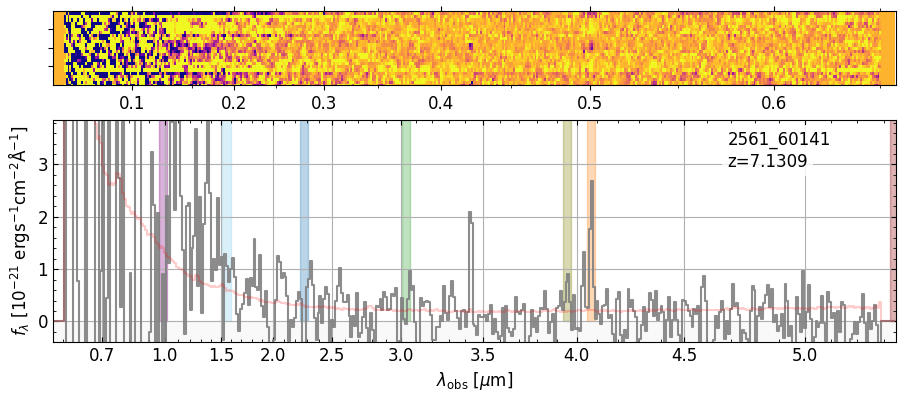}
\caption{\textit{JWST} NIRSpec/PRISM confirmation of a $z\sim7$ [OIII] candidate with UNCOVER object ID 60141 and redshift $z=7.1309$ ($\alpha=3.62035$, $\delta=-30.38860$, $\mu=1.5$, F410M $=25.4$ AB, F444W $=26.9$ AB).  The format of this figure is the same as Figure 13. The bright spectral feature at an observed wavelength of $\sim3.4 \mu$m appears to be a bad pixel artifact.}\label{spec2}
\par\end{centering}
\end{figure*}

\section{Comparison to Ly$\alpha$}\label{slya}

Studies of low-redshift analogs to cosmic dawn galaxies have shown that strong [OIII] emitters are preferentially Ly$\alpha$ emitters with an average Ly$\alpha$ to [OIII] ratio of $\sim1$ \citep{yang16,yang17}. 
This trend could be explained by density-bounded nebula which are expected to have high [OIII] flux and to have a porous ISM that facilitates the escape of Ly$\alpha$ photons \citep{nakajima13,nakajima14}. Guided by these low-redshift results, in Figure \ref{lya} we compare the [OIII] luminosity functions at $z\sim6$ and $z\sim7$ to the corresponding Ly$\alpha$ luminosity functions.  We find close agreement between the [OIII] and Ly$\alpha$ LF at $z\sim6$.  While combined Ly$\alpha$$+$[OIII]  observations are needed to confirm, this could be explained by an average Ly$\alpha$ to [OIII] ratio of $\sim1$, as expected from the low-redshift studies.

At $z\sim7$ there is no luminosity overlap between the [OIII] and Ly$\alpha$ LFs, but they both display a similar decline from their LF at $z\sim6$.  As discussed in the previous section, a larger-volume [OIII] survey is needed to constrain the bright-end of the $z\sim7$ LF and to confirm this small LF decline.  This is of interest because assuming no ISM evolution over a 165 Myr duration from $z=6$ to $7$, the LF decline, as quantified by the Ly$\alpha$ luminosity density ratio normalized by the [OIII] luminosity density ratio, can constrain the Ly$\alpha$ transmission through the IGM and the overall ionization state of the IGM at $z\sim7$.

\section{Spectroscopic Confirmations}

Using The DAWN {\it JWST} Archive, we cross-matched our $z\sim7$ [OIII] candidates with all the publicly available NIRSpec data as of fall 2023.   We found that two of our candidates have GO-2561 (UNCOVER) NIRSpec/PRISM $R=100$ spectra with UNCOVER IDs 16155 and 60141 (see Figure 13 and 14).  Both are confirmed to be $z\sim7$ [OIII] emitters, demonstrating the feasibility of our F410M-excess selection. We tentatively detect  CIII]$_{1907, 1909}$ in object 16155 similar to the Ly$\alpha$ identified galaxy at $z=7.5$ studied by \citet{hutchison19}.  The high energy required to excite CIII] typically necessitates young, low-metallicity stars and/or AGN contribution.  We defer further spectroscopic analysis to our follow-up paper.

\section{Summary}

We use publicly available UNCOVER DR1 {\it JWST}/NIRCam and {\it HST} imaging data of the strong lensing field, Abell 2744, to isolate strong (rest-frame EW$>500$\AA) [OIII]$_{5008}$ emitters at $z\sim7$ based on excess F410M flux.  To obtain accurate colors with fixed aperture photometry, we PSF-matched all bandpasses to the {\it JWST} F444 PSF.  We demonstrate that our PSF-matched F410M-selected photometry can reliably reproduce the UNCOVER DR1 cataloged F277+F356+F444W-selected photometric redshifts. We find $N=68$ $z\sim7$ [OIII] candidates with our F410M-excess search.  A large percentage of these, $22\%$, have no DR1 cataloged counterparts, showing that medium-band searches for strong emission lines can find $z\sim7$ sources that would otherwise be missed even in extremely deep
{\it JWST} broadband surveys. A subset of $N=33$ among our $z\sim7$ [OIII] candidates have deep {\it HST} veto coverage (13.68 arcmin$^2$ with F814W $5\sigma$ depth $>28$ AB). The sizes of these extreme [OIII] emitters are broadly similar to their low redshift analogs, the green peas. 

Based on the deep {\it HST} sample which is more resistant to our primary foreground contamination concern, EW$>1000$\AA\ H$\alpha$ emitters at $z\sim5$, we compute the $z\sim7$ [OIII] LF.  This work includes completeness simulations that we use to calculate the effective survey volume of the UNCOVER lensing field as a function of [OIII] luminosity.  Our LF spans a luminosity range of $41.09<\rm{log}_{10}(L/\rm{erg\,s}^{-1})<42.35$ which is an order of magnitude deeper than previous $z\sim6$ [OIII] LFs based on {\it JWST} slitless spectroscopy.  Our deep LF is well fit by a power law with a faint-end slope of $\alpha=-2.07^{+0.22}_{-0.23}$.  We compare our LF result with archival studies finding little evidence for redshift evolution from $z\sim3$--$7$. The lack of an observed LF turnover at faint luminosities suggests that the metallicities of our objects are not low enough to cause oxygen deficiency and a resulting decline in [OIII] emission. The luminosity function of [OIII] emitters matches that of Lyman-$\alpha$ at the bright end, suggesting that many [OIII] emitters are also Lyman-$\alpha$ emitters.

\noindent \begin{acknowledgments}IGBW is supported by an appointment to the NASA Postdoctoral Program at the Goddard Space Flight Center. The material is based upon work supported by NASA under award number 80GSFC21M0002. Some of the data products presented herein were retrieved from the Dawn JWST Archive (DJA). DJA is an initiative of the Cosmic Dawn Center, which is funded by the Danish National Research Foundation under grant No. 140.
\end{acknowledgments}

\bibliographystyle{aasjournal} 
\bibliography{wold}

\begin{thebibliography}{}
\expandafter\ifx\csname natexlab\endcsname\relax\def\natexlab#1{#1}\fi
\providecommand{\url}[1]{\href{#1}{#1}}
\providecommand{\dodoi}[1]{doi:~\href{http://doi.org/#1}{\nolinkurl{#1}}}
\providecommand{\doeprint}[1]{\href{http://ascl.net/#1}{\nolinkurl{http://ascl.net/#1}}}
\providecommand{\doarXiv}[1]{\href{https://arxiv.org/abs/#1}{\nolinkurl{https://arxiv.org/abs/#1}}}

\bibitem[{{Anderson} {et~al.}(2017){Anderson}, {Governato}, {Karcher}, {Quinn},
  \& {Wadsley}}]{anderson17}
{Anderson}, L., {Governato}, F., {Karcher}, M., {Quinn}, T., \& {Wadsley}, J.
  2017, \mnras, 468, 4077, \dodoi{10.1093/mnras/stx709}

\bibitem[{{Andrews} \& {Martini}(2013)}]{andrews13}
{Andrews}, B.~H., \& {Martini}, P. 2013, \apj, 765, 140,
  \dodoi{10.1088/0004-637X/765/2/140}

\bibitem[{{Atek} {et~al.}(2018){Atek}, {Richard}, {Kneib}, \&
  {Schaerer}}]{atek18}
{Atek}, H., {Richard}, J., {Kneib}, J.-P., \& {Schaerer}, D. 2018, \mnras, 479,
  5184, \dodoi{10.1093/mnras/sty1820}

\bibitem[{{Atek} {et~al.}(2015){Atek}, {Richard}, {Jauzac}, {Kneib},
  {Natarajan}, {Limousin}, {Schaerer}, {Jullo}, {Ebeling}, {Egami}, \&
  {Clement}}]{atek15}
{Atek}, H., {Richard}, J., {Jauzac}, M., {et~al.} 2015, \apj, 814, 69,
  \dodoi{10.1088/0004-637X/814/1/69}

\bibitem[{{Bertin} \& {Arnouts}(1996)}]{bertin96}
{Bertin}, E., \& {Arnouts}, S. 1996, \aaps, 117, 393

\bibitem[{{Bezanson} {et~al.}(2022){Bezanson}, {Labbe}, {Whitaker}, {Leja},
  {Price}, {Franx}, {Brammer}, {Marchesini}, {Zitrin}, {Wang}, {Weaver},
  {Furtak}, {Atek}, {Coe}, {Cutler}, {Dayal}, {van Dokkum}, {Feldmann},
  {Forster Schreiber}, {Fujimoto}, {Geha}, {Glazebrook}, {de Graaff}, {Greene},
  {Juneau}, {Kassin}, {Kriek}, {Khullar}, {Maseda}, {Mowla}, {Muzzin},
  {Nanayakkara}, {Nelson}, {Oesch}, {Pacifici}, {Pan}, {Papovich}, {Setton},
  {Shapley}, {Smit}, {Stefanon}, {Taylor}, \& {Williams}}]{bezanson22}
{Bezanson}, R., {Labbe}, I., {Whitaker}, K.~E., {et~al.} 2022, arXiv e-prints,
  arXiv:2212.04026, \dodoi{10.48550/arXiv.2212.04026}

\bibitem[{{Boucaud} {et~al.}(2016){Boucaud}, {Bocchio}, {Abergel}, {Orieux},
  {Dole}, \& {Hadj-Youcef}}]{boucaud16}
{Boucaud}, A., {Bocchio}, M., {Abergel}, A., {et~al.} 2016, \aap, 596, A63,
  \dodoi{10.1051/0004-6361/201629080}

\bibitem[{{Bouwens} {et~al.}(2022){Bouwens}, {Illingworth}, {Ellis}, {Oesch},
  \& {Stefanon}}]{bouwens22}
{Bouwens}, R.~J., {Illingworth}, G., {Ellis}, R.~S., {Oesch}, P., \&
  {Stefanon}, M. 2022, \apj, 940, 55, \dodoi{10.3847/1538-4357/ac86d1}

\bibitem[{{Bouwens} {et~al.}(2015){Bouwens}, {Illingworth}, {Oesch}, {Trenti},
  {Labb{\'e}}, {Bradley}, {Carollo}, {van Dokkum}, {Gonzalez}, {Holwerda},
  {Franx}, {Spitler}, {Smit}, \& {Magee}}]{bouwens15}
{Bouwens}, R.~J., {Illingworth}, G.~D., {Oesch}, P.~A., {et~al.} 2015, \apj,
  803, 34, \dodoi{10.1088/0004-637X/803/1/34}

\bibitem[{{Bouwens} {et~al.}(2021){Bouwens}, {Oesch}, {Stefanon},
  {Illingworth}, {Labb{\'e}}, {Reddy}, {Atek}, {Montes}, {Naidu},
  {Nanayakkara}, {Nelson}, \& {Wilkins}}]{bouwens21}
{Bouwens}, R.~J., {Oesch}, P.~A., {Stefanon}, M., {et~al.} 2021, \aj, 162, 47,
  \dodoi{10.3847/1538-3881/abf83e}

\bibitem[{{Cash}(1979)}]{cash79}
{Cash}, W. 1979, \apj, 228, 939, \dodoi{10.1086/156922}

\bibitem[{{Chary} {et~al.}(2005){Chary}, {Stern}, \& {Eisenhardt}}]{chary05}
{Chary}, R.-R., {Stern}, D., \& {Eisenhardt}, P. 2005, \apjl, 635, L5,
  \dodoi{10.1086/499205}

\bibitem[{{Cowie} {et~al.}(2011){Cowie}, {Barger}, \& {Hu}}]{cowie11}
{Cowie}, L.~L., {Barger}, A.~J., \& {Hu}, E.~M. 2011, \apj, 738, 136,
  \dodoi{10.1088/0004-637X/738/2/136}

\bibitem[{{De Barros} {et~al.}(2019){De Barros}, {Oesch}, {Labb{\'e}},
  {Stefanon}, {Gonz{\'a}lez}, {Smit}, {Bouwens}, \& {Illingworth}}]{debarros19}
{De Barros}, S., {Oesch}, P.~A., {Labb{\'e}}, I., {et~al.} 2019, \mnras, 489,
  2355, \dodoi{10.1093/mnras/stz940}

\bibitem[{{Egami} {et~al.}(2005){Egami}, {Kneib}, {Rieke}, {Ellis}, {Richard},
  {Rigby}, {Papovich}, {Stark}, {Santos}, {Huang}, {Dole}, {Le Floc'h}, \&
  {P{\'e}rez-Gonz{\'a}lez}}]{egami05}
{Egami}, E., {Kneib}, J.~P., {Rieke}, G.~H., {et~al.} 2005, \apjl, 618, L5,
  \dodoi{10.1086/427550}

\bibitem[{{Endsley} {et~al.}(2021){Endsley}, {Stark}, {Chevallard}, \&
  {Charlot}}]{endsley21}
{Endsley}, R., {Stark}, D.~P., {Chevallard}, J., \& {Charlot}, S. 2021, \mnras,
  500, 5229, \dodoi{10.1093/mnras/staa3370}

\bibitem[{{Faisst} {et~al.}(2016){Faisst}, {Capak}, {Hsieh}, {Laigle},
  {Salvato}, {Tasca}, {Cassata}, {Davidzon}, {Ilbert}, {Le F{\`e}vre},
  {Masters}, {McCracken}, {Steinhardt}, {Silverman}, {de Barros}, {Hasinger},
  \& {Scoville}}]{faisst16}
{Faisst}, A.~L., {Capak}, P., {Hsieh}, B.~C., {et~al.} 2016, \apj, 821, 122,
  \dodoi{10.3847/0004-637X/821/2/122}

\bibitem[{{Ferrarese} {et~al.}(2006){Ferrarese}, {C{\^o}t{\'e}}, {Jord{\'a}n},
  {Peng}, {Blakeslee}, {Piatek}, {Mei}, {Merritt}, {Milosavljevi{\'c}},
  {Tonry}, \& {West}}]{ferrarese06}
{Ferrarese}, L., {C{\^o}t{\'e}}, P., {Jord{\'a}n}, A., {et~al.} 2006, \apjs,
  164, 334, \dodoi{10.1086/501350}

\bibitem[{{Finkelstein} {et~al.}(2015){Finkelstein}, {Ryan}, {Papovich},
  {Dickinson}, {Song}, {Somerville}, {Ferguson}, {Salmon}, {Giavalisco},
  {Koekemoer}, {Ashby}, {Behroozi}, {Castellano}, {Dunlop}, {Faber}, {Fazio},
  {Fontana}, {Grogin}, {Hathi}, {Jaacks}, {Kocevski}, {Livermore}, {McLure},
  {Merlin}, {Mobasher}, {Newman}, {Rafelski}, {Tilvi}, \&
  {Willner}}]{finkelstein15}
{Finkelstein}, S.~L., {Ryan}, Russell~E., J., {Papovich}, C., {et~al.} 2015,
  \apj, 810, 71, \dodoi{10.1088/0004-637X/810/1/71}

\bibitem[{{Finkelstein} {et~al.}(2019){Finkelstein}, {D'Aloisio},
  {Paardekooper}, {Ryan}, {Behroozi}, {Finlator}, {Livermore}, {Upton
  Sanderbeck}, {Dalla Vecchia}, \& {Khochfar}}]{finkelstein19}
{Finkelstein}, S.~L., {D'Aloisio}, A., {Paardekooper}, J.-P., {et~al.} 2019,
  \apj, 879, 36, \dodoi{10.3847/1538-4357/ab1ea8}

\bibitem[{{Fujimoto} {et~al.}(2023){Fujimoto}, {Bezanson}, {Labbe}, {Brammer},
  {Price}, {Wang}, {Weaver}, {Fudamoto}, {Oesch}, {Williams}, {Dayal},
  {Feldmann}, {Greene}, {Leja}, {Whitaker}, {Zitrin}, {Cutler}, {Furtak},
  {Pan}, {Chemerynska}, {Kokorev}, {Miller}, {Atek}, {van Dokkum}, {Juneau},
  {Kassin}, {Khullar}, {Marchesini}, {Maseda}, {Nelson}, {Setton}, \&
  {Smit}}]{fujimoto23}
{Fujimoto}, S., {Bezanson}, R., {Labbe}, I., {et~al.} 2023, arXiv e-prints,
  arXiv:2309.07834, \dodoi{10.48550/arXiv.2309.07834}

\bibitem[{{Furtak} {et~al.}(2023){Furtak}, {Zitrin}, {Weaver}, {Atek},
  {Bezanson}, {Labb{\'e}}, {Whitaker}, {Leja}, {Price}, {Brammer}, {Wang},
  {Marchesini}, {Pan}, {Dayal}, {van Dokkum}, {Feldmann}, {Fujimoto}, {Franx},
  {Khullar}, {Nelson}, \& {Mowla}}]{furtak23}
{Furtak}, L.~J., {Zitrin}, A., {Weaver}, J.~R., {et~al.} 2023, \mnras, 523,
  4568, \dodoi{10.1093/mnras/stad1627}

\bibitem[{{Hu} {et~al.}(2021){Hu}, {Wang}, {Infante}, {Rhoads}, {Zheng},
  {Yang}, {Malhotra}, {Barrientos}, {Jiang}, {Gonz{\'a}lez-L{\'o}pez},
  {Prieto}, {Perez}, {Hibon}, {Galaz}, {Coughlin}, {Harish}, {Kong}, {Kang},
  {Khostovan}, {Pharo}, {Valdes}, {Wold}, {Walker}, \& {Zheng}}]{hu21}
{Hu}, W., {Wang}, J., {Infante}, L., {et~al.} 2021, Nature Astronomy, 5, 485,
  \dodoi{10.1038/s41550-020-01291-y}

\bibitem[{{Hutchison} {et~al.}(2019){Hutchison}, {Papovich}, {Finkelstein},
  {Dickinson}, {Jung}, {Zitrin}, {Ellis}, {Malhotra}, {Rhoads},
  {Roberts-Borsani}, {Song}, \& {Tilvi}}]{hutchison19}
{Hutchison}, T.~A., {Papovich}, C., {Finkelstein}, S.~L., {et~al.} 2019, \apj,
  879, 70, \dodoi{10.3847/1538-4357/ab22a2}

\bibitem[{{Khostovan} {et~al.}(2015){Khostovan}, {Sobral}, {Mobasher}, {Best},
  {Smail}, {Stott}, {Hemmati}, \& {Nayyeri}}]{khostovan15}
{Khostovan}, A.~A., {Sobral}, D., {Mobasher}, B., {et~al.} 2015, \mnras, 452,
  3948, \dodoi{10.1093/mnras/stv1474}

\bibitem[{{Kim} {et~al.}(2021){Kim}, {Malhotra}, {Rhoads}, \& {Yang}}]{kim21}
{Kim}, K.~J., {Malhotra}, S., {Rhoads}, J.~E., \& {Yang}, H. 2021, \apj, 914,
  2, \dodoi{10.3847/1538-4357/abf833}

\bibitem[{{Kojima} {et~al.}(2020){Kojima}, {Ouchi}, {Rauch}, {Ono}, {Nakajima},
  {Isobe}, {Fujimoto}, {Harikane}, {Hashimoto}, {Hayashi}, {Komiyama},
  {Kusakabe}, {Kim}, {Lee}, {Mukae}, {Nagao}, {Onodera}, {Shibuya}, {Sugahara},
  {Umemura}, \& {Yabe}}]{kojima20}
{Kojima}, T., {Ouchi}, M., {Rauch}, M., {et~al.} 2020, \apj, 898, 142,
  \dodoi{10.3847/1538-4357/aba047}

\bibitem[{{Konno} {et~al.}(2018){Konno}, {Ouchi}, {Shibuya}, {Ono},
  {Shimasaku}, {Taniguchi}, {Nagao}, {Kobayashi}, {Kajisawa}, {Kashikawa},
  {Inoue}, {Oguri}, {Furusawa}, {Goto}, {Harikane}, {Higuchi}, {Komiyama},
  {Kusakabe}, {Miyazaki}, {Nakajima}, \& {Wang}}]{konno18}
{Konno}, A., {Ouchi}, M., {Shibuya}, T., {et~al.} 2018, \pasj, 70, S16,
  \dodoi{10.1093/pasj/psx131}

\bibitem[{{Labb{\'e}} {et~al.}(2013){Labb{\'e}}, {Oesch}, {Bouwens},
  {Illingworth}, {Magee}, {Gonz{\'a}lez}, {Carollo}, {Franx}, {Trenti}, {van
  Dokkum}, \& {Stiavelli}}]{Labbe13}
{Labb{\'e}}, I., {Oesch}, P.~A., {Bouwens}, R.~J., {et~al.} 2013, \apjl, 777,
  L19, \dodoi{10.1088/2041-8205/777/2/L19}

\bibitem[{{Llerena} {et~al.}(2024){Llerena}, {Amor{\'\i}n}, {Pentericci},
  {Arrabal Haro}, {Backhaus}, {Bagley}, {Calabr{\`o}}, {Cleri}, {Davis},
  {Dickinson}, {Finkelstein}, {Gawiser}, {Grogin}, {Hathi}, {Hirschmann},
  {Kartaltepe}, {Koekemoer}, {McGrath}, {Mobasher}, {Napolitano}, {Papovich},
  {Pirzkal}, {Trump}, {Wilkins}, \& {Yung}}]{llerena24}
{Llerena}, M., {Amor{\'\i}n}, R., {Pentericci}, L., {et~al.} 2024, arXiv
  e-prints, arXiv:2403.05362, \dodoi{10.48550/arXiv.2403.05362}

\bibitem[{{Matthee} {et~al.}(2023){Matthee}, {Mackenzie}, {Simcoe}, {Kashino},
  {Lilly}, {Bordoloi}, \& {Eilers}}]{matthee23}
{Matthee}, J., {Mackenzie}, R., {Simcoe}, R.~A., {et~al.} 2023, \apj, 950, 67,
  \dodoi{10.3847/1538-4357/acc846}

\bibitem[{{McLinden} {et~al.}(2011){McLinden}, {Finkelstein}, {Rhoads},
  {Malhotra}, {Hibon}, {Richardson}, {Cresci}, {Quirrenbach}, {Pasquali},
  {Bian}, {Fan}, \& {Woodward}}]{mclinden11}
{McLinden}, E.~M., {Finkelstein}, S.~L., {Rhoads}, J.~E., {et~al.} 2011, \apj,
  730, 136, \dodoi{10.1088/0004-637X/730/2/136}

\bibitem[{{Nakajima} \& {Ouchi}(2014)}]{nakajima14}
{Nakajima}, K., \& {Ouchi}, M. 2014, \mnras, 442, 900,
  \dodoi{10.1093/mnras/stu902}

\bibitem[{{Nakajima} {et~al.}(2013){Nakajima}, {Ouchi}, {Shimasaku},
  {Hashimoto}, {Ono}, \& {Lee}}]{nakajima13}
{Nakajima}, K., {Ouchi}, M., {Shimasaku}, K., {et~al.} 2013, \apj, 769, 3,
  \dodoi{10.1088/0004-637X/769/1/3}

\bibitem[{{Ning} {et~al.}(2022){Ning}, {Jiang}, {Zheng}, \& {Wu}}]{ning22}
{Ning}, Y., {Jiang}, L., {Zheng}, Z.-Y., \& {Wu}, J. 2022, \apj, 926, 230,
  \dodoi{10.3847/1538-4357/ac4268}

\bibitem[{{Rhoads} {et~al.}(2023){Rhoads}, {Wold}, {Harish}, {Kim}, {Pharo},
  {Malhotra}, {Gabrielpillai}, {Jiang}, \& {Yang}}]{rhoads23}
{Rhoads}, J.~E., {Wold}, I. G.~B., {Harish}, S., {et~al.} 2023, \apjl, 942,
  L14, \dodoi{10.3847/2041-8213/acaaaf}

\bibitem[{{Richard} {et~al.}(2021){Richard}, {Claeyssens}, {Lagattuta},
  {Guaita}, {Bauer}, {Pello}, {Carton}, {Bacon}, {Soucail}, {Lyon}, {Kneib},
  {Mahler}, {Cl{\'e}ment}, {Mercier}, {Variu}, {Tamone}, {Ebeling}, {Schmidt},
  {Nanayakkara}, {Maseda}, {Weilbacher}, {Bouch{\'e}}, {Bouwens}, {Wisotzki},
  {de la Vieuville}, {Martinez}, \& {Patr{\'\i}cio}}]{richard21}
{Richard}, J., {Claeyssens}, A., {Lagattuta}, D., {et~al.} 2021, \aap, 646,
  A83, \dodoi{10.1051/0004-6361/202039462}

\bibitem[{{Rinaldi} {et~al.}(2023){Rinaldi}, {Caputi}, {Costantin}, {Gillman},
  {Iani}, {P{\'e}rez-Gonz{\'a}lez}, {{\"O}stlin}, {Colina}, {Greve},
  {Noorgard-Nielsen}, {Wright}, {Alonso-Herrero}, {{\'A}lvarez-M{\'a}rquez},
  {Eckart}, {Garc{\'\i}a-Mar{\'\i}n}, {Hjorth}, {Ilbert}, {Kendrew}, {Labiano},
  {Le F{\`e}vre}, {Pye}, {Tikkanen}, {Walter}, {van der Werf}, {Ward},
  {Annunziatella}, {Azzollini}, {Bik}, {Boogaard}, {Bosman}, {Crespo
  G{\'o}mez}, {Jermann}, {Langeroodi}, {Melinder}, {Meyer}, {Moutard},
  {Peissker}, {Topinka}, {van Dishoeck}, {G{\"u}del}, {Henning}, {Lagage},
  {Ray}, {Vandenbussche}, {Waelkens}, {Navarro-Carrera}, \&
  {Kokorev}}]{Rinaldi23}
{Rinaldi}, P., {Caputi}, K.~I., {Costantin}, L., {et~al.} 2023, \apj, 952, 143,
  \dodoi{10.3847/1538-4357/acdc27}

\bibitem[{{Roberts-Borsani} {et~al.}(2016){Roberts-Borsani}, {Bouwens},
  {Oesch}, {Labbe}, {Smit}, {Illingworth}, {van Dokkum}, {Holden}, {Gonzalez},
  {Stefanon}, {Holwerda}, \& {Wilkins}}]{Roberts-Borsani16}
{Roberts-Borsani}, G.~W., {Bouwens}, R.~J., {Oesch}, P.~A., {et~al.} 2016,
  \apj, 823, 143, \dodoi{10.3847/0004-637X/823/2/143}

\bibitem[{{Schaerer} \& {de Barros}(2009)}]{Schaerer09}
{Schaerer}, D., \& {de Barros}, S. 2009, \aap, 502, 423,
  \dodoi{10.1051/0004-6361/200911781}

\bibitem[{{Schechter}(1976)}]{schechter76}
{Schechter}, P. 1976, \apj, 203, 297, \dodoi{10.1086/154079}

\bibitem[{{Shen} {et~al.}(2020){Shen}, {Vogelsberger}, {Nelson}, {Pillepich},
  {Tacchella}, {Marinacci}, {Torrey}, {Hernquist}, \& {Springel}}]{shen20}
{Shen}, X., {Vogelsberger}, M., {Nelson}, D., {et~al.} 2020, \mnras, 495, 4747,
  \dodoi{10.1093/mnras/staa1423}

\bibitem[{{Smit} {et~al.}(2014){Smit}, {Bouwens}, {Labb{\'e}}, {Zheng},
  {Bradley}, {Donahue}, {Lemze}, {Moustakas}, {Umetsu}, {Zitrin}, {Coe},
  {Postman}, {Gonzalez}, {Bartelmann}, {Ben{\'\i}tez}, {Broadhurst}, {Ford},
  {Grillo}, {Infante}, {Jimenez-Teja}, {Jouvel}, {Kelson}, {Lahav}, {Maoz},
  {Medezinski}, {Melchior}, {Meneghetti}, {Merten}, {Molino}, {Moustakas},
  {Nonino}, {Rosati}, \& {Seitz}}]{Smit14}
{Smit}, R., {Bouwens}, R.~J., {Labb{\'e}}, I., {et~al.} 2014, \apj, 784, 58,
  \dodoi{10.1088/0004-637X/784/1/58}

\bibitem[{{Smit} {et~al.}(2015){Smit}, {Bouwens}, {Franx}, {Oesch}, {Ashby},
  {Willner}, {Labb{\'e}}, {Holwerda}, {Fazio}, \& {Huang}}]{smit15}
{Smit}, R., {Bouwens}, R.~J., {Franx}, M., {et~al.} 2015, \apj, 801, 122,
  \dodoi{10.1088/0004-637X/801/2/122}

\bibitem[{{Stark} {et~al.}(2013){Stark}, {Schenker}, {Ellis}, {Robertson},
  {McLure}, \& {Dunlop}}]{stark13}
{Stark}, D.~P., {Schenker}, M.~A., {Ellis}, R., {et~al.} 2013, \apj, 763, 129,
  \dodoi{10.1088/0004-637X/763/2/129}

\bibitem[{{Suess} {et~al.}(2024){Suess}, {Weaver}, {Price}, {Pan}, {Wang},
  {Bezanson}, {Brammer}, {Cutler}, {Labbe}, {Leja}, {Williams}, {Whitaker},
  {Dayal}, {de Graaff}, {Feldmann}, {Franx}, {Fudamoto}, {Fujimoto}, {Furtak},
  {Goulding}, {Greene}, {Khullar}, {Kokorev}, {Kriek}, {Lorenz}, {Marchesini},
  {Maseda}, {Matthee}, {Miller}, {Mitsuhashi}, {Mowla}, {Muzzin}, {Naidu},
  {Nanayakkara}, {Nelson}, {Oesch}, {Setton}, {Shipley}, {Smit}, {Spilker},
  {van Dokkum}, \& {Zitrin}}]{suess24}
{Suess}, K.~A., {Weaver}, J.~R., {Price}, S.~H., {et~al.} 2024, arXiv e-prints,
  arXiv:2404.13132, \dodoi{10.48550/arXiv.2404.13132}

\bibitem[{{Sun} {et~al.}(2023){Sun}, {Egami}, {Pirzkal}, {Rieke}, {Baum},
  {Boyer}, {Boyett}, {Bunker}, {Cameron}, {Curti}, {Eisenstein}, {Gennaro},
  {Greene}, {Jaffe}, {Kelly}, {Koekemoer}, {Kumari}, {Maiolino}, {Maseda},
  {Perna}, {Rest}, {Robertson}, {Schlawin}, {Smit}, {Stansberry}, {Sunnquist},
  {Tacchella}, {Williams}, \& {Willmer}}]{sun23}
{Sun}, F., {Egami}, E., {Pirzkal}, N., {et~al.} 2023, \apj, 953, 53,
  \dodoi{10.3847/1538-4357/acd53c}

\bibitem[{{Taylor} {et~al.}(2020){Taylor}, {Barger}, {Cowie}, {Hu}, \&
  {Songaila}}]{taylor20}
{Taylor}, A.~J., {Barger}, A.~J., {Cowie}, L.~L., {Hu}, E.~M., \& {Songaila},
  A. 2020, \apj, 895, 132, \dodoi{10.3847/1538-4357/ab8ada}

\bibitem[{{Taylor} {et~al.}(2021){Taylor}, {Cowie}, {Barger}, {Hu}, \&
  {Songaila}}]{taylor21}
{Taylor}, A.~J., {Cowie}, L.~L., {Barger}, A.~J., {Hu}, E.~M., \& {Songaila},
  A. 2021, \apj, 914, 79, \dodoi{10.3847/1538-4357/abfc4b}

\bibitem[{{Tremonti} {et~al.}(2004){Tremonti}, {Heckman}, {Kauffmann},
  {Brinchmann}, {Charlot}, {White}, {Seibert}, {Peng}, {Schlegel}, {Uomoto},
  {Fukugita}, \& {Brinkmann}}]{tremonti04}
{Tremonti}, C.~A., {Heckman}, T.~M., {Kauffmann}, G., {et~al.} 2004, \apj, 613,
  898, \dodoi{10.1086/423264}

\bibitem[{{Treu} {et~al.}(2015){Treu}, {Schmidt}, {Brammer}, {Vulcani}, {Wang},
  {Brada{\v{c}}}, {Dijkstra}, {Dressler}, {Fontana}, {Gavazzi}, {Henry},
  {Hoag}, {Huang}, {Jones}, {Kelly}, {Malkan}, {Mason}, {Pentericci},
  {Poggianti}, {Stiavelli}, {Trenti}, \& {von der Linden}}]{treu15}
{Treu}, T., {Schmidt}, K.~B., {Brammer}, G.~B., {et~al.} 2015, \apj, 812, 114,
  \dodoi{10.1088/0004-637X/812/2/114}

\bibitem[{{Wang} {et~al.}(2024){Wang}, {Leja}, {Labb{\'e}}, {Bezanson},
  {Whitaker}, {Brammer}, {Furtak}, {Weaver}, {Price}, {Zitrin}, {Atek}, {Coe},
  {Cutler}, {Dayal}, {van Dokkum}, {Feldmann}, {Marchesini}, {Franx},
  {F{\"o}rster Schreiber}, {Fujimoto}, {Geha}, {Glazebrook}, {de Graaff},
  {Greene}, {Juneau}, {Kassin}, {Kriek}, {Khullar}, {Maseda}, {Mowla},
  {Muzzin}, {Nanayakkara}, {Nelson}, {Oesch}, {Pacifici}, {Pan}, {Papovich},
  {Setton}, {Shapley}, {Smit}, {Stefanon}, {Suess}, {Taylor}, \&
  {Williams}}]{wang24}
{Wang}, B., {Leja}, J., {Labb{\'e}}, I., {et~al.} 2024, \apjs, 270, 12,
  \dodoi{10.3847/1538-4365/ad0846}

\bibitem[{{Weaver} {et~al.}(2024){Weaver}, {Cutler}, {Pan}, {Whitaker},
  {Labb{\'e}}, {Price}, {Bezanson}, {Brammer}, {Marchesini}, {Leja}, {Wang},
  {Furtak}, {Zitrin}, {Atek}, {Chemerynska}, {Coe}, {Dayal}, {van Dokkum},
  {Feldmann}, {F{\"o}rster Schreiber}, {Franx}, {Fujimoto}, {Fudamoto},
  {Glazebrook}, {de Graaff}, {Greene}, {Juneau}, {Kassin}, {Kriek}, {Khullar},
  {Maseda}, {Mowla}, {Muzzin}, {Nanayakkara}, {Nelson}, {Oesch}, {Pacifici},
  {Papovich}, {Setton}, {Shapley}, {Shipley}, {Smit}, {Stefanon}, {Taylor},
  {Weibel}, \& {Williams}}]{weaver24}
{Weaver}, J.~R., {Cutler}, S.~E., {Pan}, R., {et~al.} 2024, \apjs, 270, 7,
  \dodoi{10.3847/1538-4365/ad07e0}

\bibitem[{{Wilkins} {et~al.}(2023){Wilkins}, {Lovell}, {Vijayan}, {Irodotou},
  {Adams}, {Roper}, {Caruana}, {Matthee}, {Seeyave}, {Conselice},
  {P{\'e}rez-Gonz{\'a}lez}, {Turner}, {Donnellan}, {Verma}, \&
  {Trussler}}]{wilkins23}
{Wilkins}, S.~M., {Lovell}, C.~C., {Vijayan}, A.~P., {et~al.} 2023, \mnras,
  522, 4014, \dodoi{10.1093/mnras/stad1126}

\bibitem[{{Withers} {et~al.}(2023){Withers}, {Muzzin}, {Ravindranath},
  {Sarrouh}, {Abraham}, {Asada}, {Bradac}, {Brammer}, {Desprez}, {Iyer},
  {Martis}, {Mowla}, {Noirot}, {Sawicki}, {Strait}, \& {Willott}}]{Withers23}
{Withers}, S., {Muzzin}, A., {Ravindranath}, S., {et~al.} 2023, arXiv e-prints,
  arXiv:2304.11181, \dodoi{10.48550/arXiv.2304.11181}

\bibitem[{{Yang} {et~al.}(2016){Yang}, {Malhotra}, {Gronke}, {Rhoads},
  {Dijkstra}, {Jaskot}, {Zheng}, \& {Wang}}]{yang16}
{Yang}, H., {Malhotra}, S., {Gronke}, M., {et~al.} 2016, \apj, 820, 130,
  \dodoi{10.3847/0004-637X/820/2/130}

\bibitem[{{Yang} {et~al.}(2017){Yang}, {Malhotra}, {Gronke}, {Rhoads},
  {Leitherer}, {Wofford}, {Jiang}, {Dijkstra}, {Tilvi}, \& {Wang}}]{yang17}
---. 2017, \apj, 844, 171, \dodoi{10.3847/1538-4357/aa7d4d}

\end{thebibliography}

\end{document}